\documentclass[aps,prb,showpacs,twocolumn,amsmath,amssymb]{revtex4-1}
\usepackage{graphicx}
\usepackage{epstopdf}
\usepackage{bm}% bold math
\usepackage{subfigure}
\usepackage{sidecap}

\usepackage{graphicx}
\usepackage{color}
\usepackage{epstopdf}
\usepackage{amssymb}
\usepackage{amsmath}

\usepackage{bm}% bold math
\graphicspath{{Figures/}}
\usepackage{subfigure}
\usepackage{sidecap}
\usepackage {times}

\newcommand{\al}{\alpha}

\newcommand{\De}{\Delta}

\newcommand{\be}{\begin{equation}}
\newcommand{\ee}{\end{equation}}

\newcommand{\bea}{\begin{eqnarray}}
\newcommand{\eea}{\end{eqnarray}}
\newcommand{\bd}{\begin{displaymath}}
\newcommand{\ed}{\end{displaymath}}
\newcommand{\ba}{\begin{array}}
\newcommand{\ea}{\end{array}}
\newcommand{\bi}{\begin{itemize}}
\newcommand{\ei}{\end{itemize}}
\newcommand{\bc}{\begin{center}}
\newcommand{\ec}{\end{center}}
\newcommand{\bfl}{\begin{flushleft}}
\newcommand{\efl}{\end{flushleft}}
\newcommand{\bfr}{\begin{flushright}}
\newcommand{\efr}{\end{flushright}}
\newcommand{\non}{\nonumber}

\newcommand{\bl}{\begin{aligned}}
\newcommand{\el}{\end{aligned}}

\newcommand{\bJ}{\bar{J}}

\newcommand{\tl}{\tilde{\lambda}}

\newcommand{\tal}{\tilde{\alpha}}

\newcommand{\tsh}{\tilde{s}}
\newcommand{\tch}{\tilde{c}}

\newcommand{\tDe}{\tilde{\Delta}}

\newcommand{\fs}{\frac{1}{2}}

\newcommand{\beh}{\frac{\beta}{2}}

\newcommand{\om}{i\omega_n}

\newcommand{\ra}{\rangle}

\newcommand{\UR}{URu$_2$Si$_2$}

\def\ket#1{\left\vert #1 \right\rangle}

\def\bR{{\bf R}} 
\def\bk{{\bf k}}  \def\bq{{\bf q}}

\def\bQ{{\bf Q}}   
  
 \def\bd{{\bf d}}  \def\bJ{{\bf J}}

 \def\bJ{{\bf J}}

\def\bra{\langle}
\def\ket{\rangle}
\def\={\!\!\!&=&\!\!\!}
\def\+{\!\!\!&&\!\!\!+~}
\def\-{\!\!\!&&\!\!\!-~}

\usepackage{color}
\usepackage[dvipsnames]{xcolor}
\usepackage{xcolor}
\usepackage[colorlinks=true,citecolor=blue]{hyperref}

\usepackage{hyperref,cleveref}
\begin{document}

\title{Induced order and collective excitations in three-singlet quantum magnets}

\author{Peter Thalmeier}
\affiliation{Max Planck Institute for the  Chemical Physics of Solids, D-01187 Dresden, Germany}
\date{\today}

\begin{abstract}
The quantum magnetism in a three-singlet model (TSM) with singlet crystalline electric field (CEF) states interacting on a lattice is investigated, motivated by  its appearance in compounds with $4f^2$ and $5f^2$ electronic structure. Contrary to
conventional (semi-classical) magnetism there are no preformed moments above the ordering temperature $T_m$.
They appear spontaneously as induced or excitonic moments due to singlet-singlet mixing at $T_m$. In most cases the transition is
of second order, however for large matrix elements between the excited states it turns into a first order transition at a
critical point. Furthermore we derive the excitonic mode spectrum and its quantum critical soft mode behaviour which
 leads to the criticality condition for induced order as expressed 
in terms of the control parameters of the TSM and discuss the distinctions to the previously known two-singlet case. We also derive the
temperature dependence of order parameters for second and first order transitions  and the exciton spectrum in the induced magnetic phase. \end{abstract}

%\pacs{ }

\maketitle
%%%%%%%%%%%%%%%%%%%%%%%%%%%%%%%%%%%%%%%%%%%%%%%%%%%%%%

\section{Introduction}
\label{sec:introduction}

In ordinary (semi-classical) magnets the individual magnetic moments at every lattice site exist already above
the ordering temperature $T_m$ \cite{majlis:07}. This holds even in strongly frustrated local-moment systems which may have a vanishing ordering temperature when fine-tuned to a spin-liquid regime where quantum fluctuations destroy the moment of the ground state but nevertheless the Curie-Weiss signature of local moments remains for elevated temperatures  \cite{schmidt:17a,schmidt:17b,schmidt:15}. There are, however, true quantum magnets which do not have freely rotating magnetic moments above $T_m$ in the semi-classical sense as witnessed by an absence of the Curie-Weiss type susceptibility for some region above $T_m$. In these compounds with partly filled 4f or 5f electron shells the degenerate ground state with integer (non-Kramers) total angular momentum J, created by spin-orbit coupling splits due to the local crystalline electric field (CEF) into a series of multiplets \cite{jensen:91} . They  belong to irreducible representatations $\Gamma_i$  which may comprise singlets, doublets or triplets depending on the symmetry of the CEF and the concrete CEF potential. For tetragonal or lower symmetry it is possible that the ground state and lowest excited states are all singlets without magnetic moment meaning $\langle\Gamma_i |\bJ |\Gamma_i\rangle = 0$. Nevertheless magnetic order occurs below the transition temperature $T_m$. This order cannot be interpreted in the usual semiclassical way as an alignment of preexisting  moments which then have collective semi-classical spin wave excitations as Goldstone modes. In the latter case quantum effects enter only through the possible reduction of the saturation moment due to zero point fluctuations leading to spin wave contribution to the ground state energy.\\

For the CEF systems with split singlet low lying states  the local  moments instead appear only simultaneously with the magnetic order as a true quantum effect due to the mixing of singlet states caused by inter-site exchange interactions. This 'induced' or 'excitonic' magnetic order has been observed primarily in various Pr (4f$^2$) and U (5f$^2$) compounds with two f- electrons which lead to CEF schemes with singlet ground state and possibly also low energy excited singlets. However it can also be found in f-electron compounds with higher even f-occupation, like e.g. Tb ($4f^8$).  In the cubic $(O_h)$ symmetry cases with singlet ground state the excited states must be degenerate as in fcc Pr \cite{birgeneau:72,cooper:72}, PrSb \cite{mcwhan:79}, Pr$_3$Tl \cite{buyers:75} and TbSb \cite{holden:74} (singlet - triplet).   Examples with hexagonal $(D_{6h})$  structure are metallic Pr (singlet-doublet) \cite{bak:75,houmann:79,jensen:87,jensen:91} and UPd$_2$Al$_3$ \cite{thalmeier:02} (singlet-singlet). Tetragonal $(D_{4h})$ cases are Pr$_2$CuO$_4$ \cite{sumarlin:95} (singlet-doublet) and URu$_2$Si$_2$ \cite{broholm:91,santini:94,kusunose:11} (three singlets).  The lower the symmetry the more likely one can have multiple low-energy singlets.
The most promising class in this respect has orthorhombic symmetry $(D_{2h}, D_{2})$ which has only singlets left as in PrCu$_2$ \cite{kawarazaki:95,naka:05}, PrNi \cite{tiden:06,savchenkov:19}, Tb$_3$Ga$_5$O$_{12}$ \cite{wawr:19} and Pr$_5$Ge$_4$ \cite{rao:04}. In the U-compounds, however, the situation may be more complicated due to only partial localisation of 5f-electrons \cite{zwicknagl:03,haule:09}. Since there is no degeneracy in the local 4f or 5f basis states there can also be no continuous symmetry for the exchange Hamiltonian. Therefore the collective excitations in the ordered phase may not be interpreted as spin waves resulting from coupled local spin precessions but rather as dispersive singlet-singlet (or singlet-doublet and singlet-triplet) excitation modes due to intersite exchange, commonly termed 'magnetic excitons'. These  are already present above the ordering temperature. The ordering  is characterised by a softening of one of these modes at $T_m$ and a subsequent stiffening again further below.\\

This type of excitonic magnetism has been considered analytically primarily within the two-singlet model \cite{grover:65,wang:68,jensen:91,thalmeier:02}. A fully numerical treatment for a multilevel CEF-system is also possible
\cite{rotter:12}. However, for a deeper understanding of induced excitonic moment ordering and their finite temperture properties analytical investigations are desirable. In particular the influence of physical parameters like splittings, nondiagonal matrix elements and exchange which define dimensionless control parameters on the transition temperature, saturation moment and mode softening are rendered understandable only when explicit analytical expressions can be derived. This becomes quite involved beyond the two-singlet model. The latter is, however, an over-simplificiation as very often more levels, in particular another singlet state are present, as ,e.g,. in PrNi, PrCu$_2$  and URu$_2$Si$_2$. \\

Therefore in this work we give a detailed analytical treatment of induced moment behaviour in the physically important three-singlet model (TSM) relevant for non-Kramers f-electron systems in lower than cubic symmetry, in particular we investigate the case of orthorhombic symmetry. We will focus on the mode spectrum, transition temperature and saturation moment and how they are influenced by the larger set of control parameters of this extended model. We show that under suitable conditons temperature variation induces hybridization of exciton modes in addition to the changes of intensity pattern. Furthermore we derive an algebraic equation that completely determines the transition temperature for the effective two control parameters and arbitrary splitting ratio of the TSM. In the symmetric TSM explicit closed expressions for T$_m$ are presented. Furthermore we give a comparative treatment of the exciton mode dispersions within random phase approximation (RPA) response function formalism and Bogoliubov quasiparticle picture and show that they give largely equivalent results, also for the phase boundary between disordered and excitonic phase. Finally within the RPA formalism we will investigate the change of mode dispersions and intensity in the induced moment phase. This work is mainly theoretically motivated with the aim to analyze and understand the three-singlet model and its significance for excitonic magnetism in detail.

\section{The three singlet model}
\label{sec:model}

\begin{figure}
\vspace{1cm}
\includegraphics[width=0.90\columnwidth]{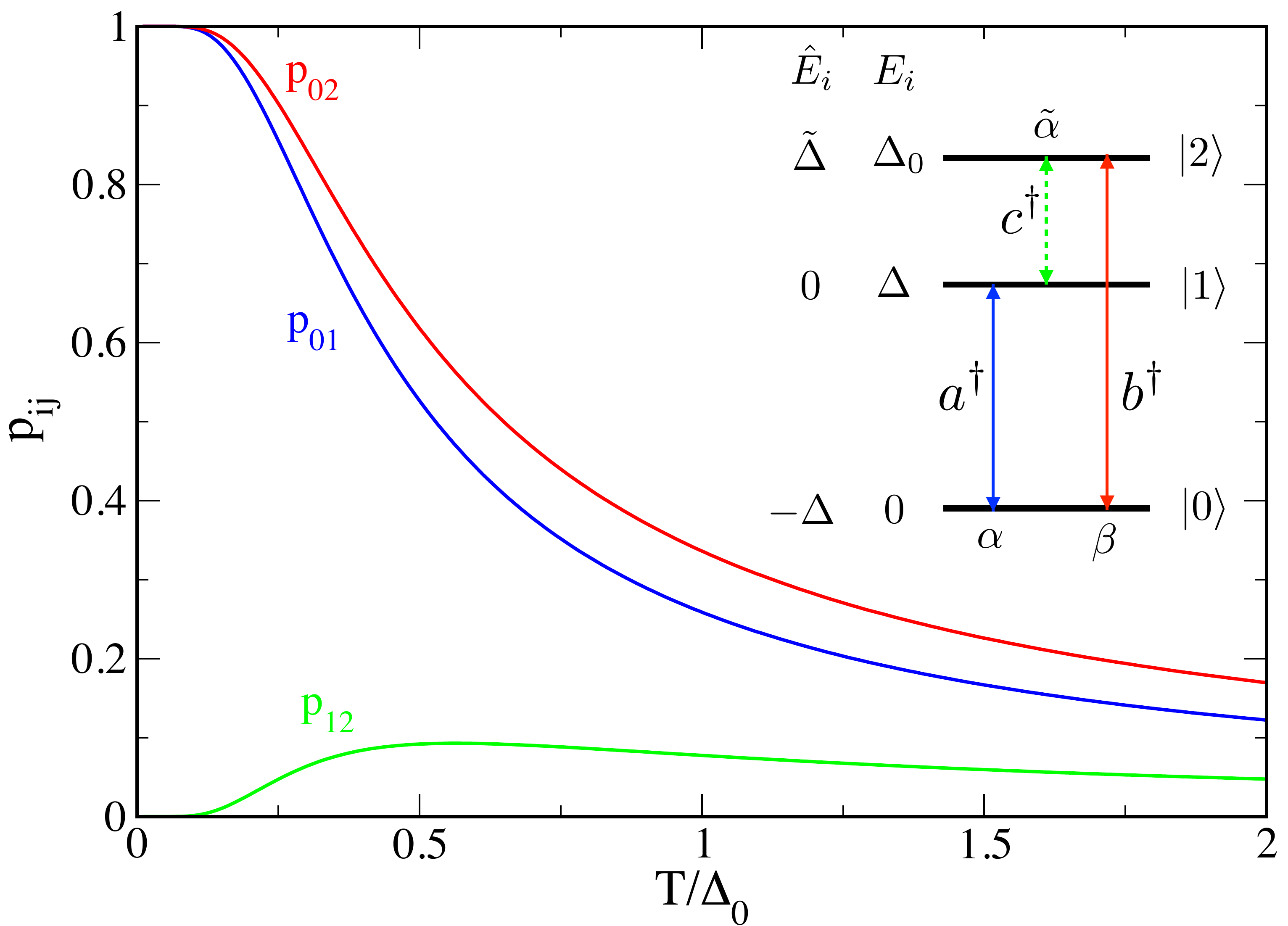}
\caption{Inset: Designations of the general TSM with singlet states $|i\rangle$ $(i=1-3)$ and respective energies 
$E_i$ or $\hat{E}_i=E_i-\Delta$, here $\tDe=\Delta_0-\Delta$. 
Matrix elements of $J_z$ are denoted by $\alpha,\beta,\tal$ (one of them must vanish in the nonmagnetic state) and 
boson excitation operators by $a^\dagger, b^\dagger, c^\dagger$. The special fully symmetric TSM is defined by $r=\tDe/\Delta=1$ and $\tal=\alpha$. Transition arrows correspond to the thermal occupation differences $p_{ij}$ shown in the main figure. Here $r=0.5$, the energy scale is $\Delta_0$ in all figures.}
\label{fig:occ-level}
\end{figure}

We keep the specifications  of the three-singlet model (TSM) illustrated in the inset of Fig.~\ref{fig:occ-level} as general as possible,
as far as splittings and magnetic matrix elements of $J_z$ are concerned. However, having orthorhombic CEF system in mind  the latter are assumed to be of uniaxial character due to $\langle\Gamma_i |J_{x,y} |\Gamma_j\rangle = 0$ (Sec.~\ref{subsect:ortho}, Eq.~(\ref{eqn:Jzmat})). The three singlets are denoted by $|i\rangle$ $(i=1-3)$ with increasing level energies $E_i=0,\Delta,\Delta_0$,
or shifted energies $\hat{E}_i=E_i-\Delta= -\Delta,0,\tDe$ which are more convenient for finite-temperature properties; here we defined $\tDe=\Delta_0-\Delta$. The CEF Hamiltonian in can be written in terms of standard basis operators $L_{ij}=|i\rangle\langle j|$ as 
\bea
H_{CEF}=\sum_i E_i|i\rangle\langle i|
\label{eqn:CEFham}
\eea
The  total angular momentum component $J_z$ in this representation is given by $J_z=\sum_{ij}\langle i|J_z|j\rangle L_{ij}$. Without restriction this leaves us with three possible independent matrix elements $\alpha=\langle 0|J_z|1\rangle $, $\beta=\langle 0|J_z|2\rangle $ and   $\tal=\langle 1|J_z|2\rangle $ (Sec.~\ref{subsect:ortho}). The latter plays only a role at finite temperature $T$ when excited states are populated with population numbers $p_i=Z^{-1}\exp(-E_i/T)$ where $Z=\sum_j\exp(-E_j/T)$ is the three-singlet partition function.
The $N_f=3$ CEF wave functions may each be gauged by an arbitrary phase factor $\exp(i\phi_n)$ $(n=1..N_f)$. Furthermore there are $\fs N_f(N_f-1)=3$ excitation matrix elements between those states. Therefore in the TSM all matrix elements $\alpha,\beta,\tal$ may be chosen as
real without loss of generality. We will pay particular attention to the special case of the (fully) symmetric three singlet model which is defined by $\tDe=\Delta; \tal=\alpha$ in Fig.~\ref{fig:occ-level}. The magnetic properties of the model are characterized by the three possible dimensionless control parameters
\bea
\xi_\alpha=\frac{2\alpha^2I_e}{\Delta};\;\;\;
\xi_\beta=\frac{2\beta^2I_e}{\Delta_0};\;\;\;
\xi_{\tal}=\frac{2\tal^2I_e}{\tDe}
\label{eqn:control3}
\eea
which characterize the intersite-coupling strenghts of the three transitions with $I_e(\bq)$ denoting the Fourier transform of the  inter-site exchange in Eqs.~(\ref{eqn:hex},\ref{eqn:iex}) at the wave vector of incipient induced order where it is at maximum value. It may be $\bq=0$ ferromagnetic (FM) , general incommensurate or $\bq=\bQ=(\pi,\pi,\pi)$ antiferromagnetic (AF). In the following we focus on the latter case.
As we shall see now in the paramagnetic phase one of the three matrix elements or control parameters must vanish due to the requirements of time reversal symmetry. This leaves us with the three possible cases of TSM's depicted in Fig.~\ref{fig:TSM3}.

%
% %%%%%%%%%%%%%%%%%%%%% figure %%%%%%%%%%%%%%%%%%%%%%%%%%%%
%%%%%%%%%%%%%%%%%%%%%%fig%%%%%%%%%%%%%%%%%%%%%%%%%%%%%%%
%

\subsection{The orthorhombic three singlet model}
\label{subsect:ortho}

To realize the general TSM in a concrete CEF for given $J$ is not so straightforward as it may seem.
This is connected with the angular momentum structure of CEF eigenstates  and their  behaviour under time reversal.
As already indicated in the introduction cubic symmetry does not allow the TSM. In tetragonal $D_{4h}$ symmetry
the TSM is only realized in one specific form (Fig.~\ref{fig:TSM3}(c)) (see discussion in Appendix \ref{sect:app0}). 
%
% %%%%%%%%%%%%%%%%%%%%% figure %%%%%%%%%%%%%%%%%%%%%%%%%%%%
\begin{figure}
%\vspace{1cm}
\includegraphics[width=0.99\columnwidth]{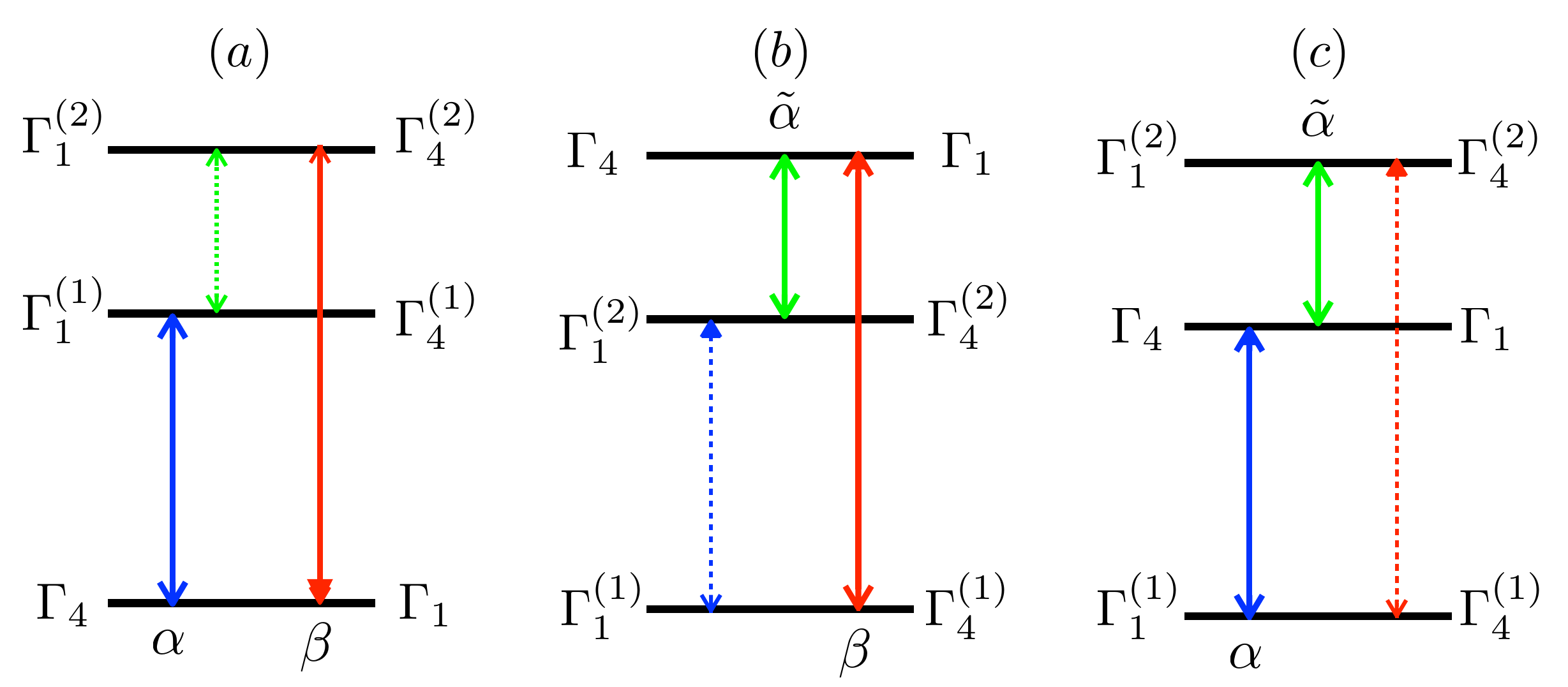}
\caption{Three possible arrangements of dipolar $(J_z)$ transition matrix elements $(\alpha,\beta,\tal)$ in the orthorhombic TSM
consisting of $(\Gamma_1,\Gamma_4^{(1,2}))$ or $(\Gamma_4,\Gamma_1^{(1,2}))$   representations. In the paramagnetic phase 
only two elements (thick lines) can be
nonzero due to $\Theta$-invariance. Transitions with dashed lines vanish because they connect states with equal $\Theta$; they are
however induced and nonzero in the magnetic state. Similar diagrams hold for the representation pair $(\Gamma_2,\Gamma_3)$ 
(cf. Eq.~(\ref{eqn:Jzmat})). There always have to be two (inequivalent) representations of the same type in the TSM for the $J_z$ structure presented.}
\label{fig:TSM3}
\end{figure}
%%%%%%%%%%%%%%%%%%%%%%fig%%%%%%%%%%%%%%%%%%%%%%%%%%%%%%%
%
Therefore we relax  to orthorhombic symmetry $D_{2h}$ where all the CEF states have to be singlet $\Gamma_i$, (i=1-4)
representations and the TSM is naturally possible. As mentioned before there are several physical realisations in the orthorhombic symmetry class. The decomposition e.g. for $J=4$ leads to nine (3$\Gamma_1\oplus 2\Gamma_2\oplus 2\Gamma_3\oplus 2\Gamma_4$) singlets.
They may be grouped according to their behaviour under time reversal symmetry operation $\Theta$  \cite{abragam:70}. For a CEF state $|\psi\ket$ written as linear combination $|\psi\ket=\sum_M c_M|M\ket$ $(|M|\leq J)$ for a given J the action of $\Theta$ is defined as 
$|\psi\ket^K=\Theta |\psi\ket=\sum_M c^*_M(-1)^{J-M}|-M\ket$. The orthorhombic singlets for $J=4$ may be expressed as \cite{wawr:19} linear combinations of $|M\ket_\pm=|M\ket\pm|-M\ket$ with only real coefficients according to
\bea
|\Gamma_1\ket&=&\sum_{M=0,2,4}b_M|M\ket_+;\;\; |\Gamma_2\ket=\sum_{M=1,3}a'_M|M\ket_- \nonumber\\
|\Gamma_3\ket&=&\sum_{M=1,3}a_M|M\ket_+;\;\; |\Gamma_4\ket=\sum_{M=2,4}b'_M|M\ket_-
\eea
This means that $|\Gamma_{1,2}\ket^K=|\Gamma_{1,2}\ket$ are even $\Theta=1$) and $|\Gamma_{3,4}\ket^K=-|\Gamma_{3,4}\ket$ are odd $(\Theta=-1)$  under time reversal. Because $J_z$ is also odd it has matrix elements only among singlets with {\it opposite} $\Theta$.
From those only two are different from zero:
\bea
\bra\Gamma_1|J_z|\Gamma_4\ket&=&2\sum_{M=2,4}Mb_Mb'_M\nonumber\\
\bra\Gamma_2|J_z|\Gamma_3\ket&=&2\sum_{M=1,3}Ma_Ma'_M
\label{eqn:Jzmat}
\eea
At the same time one observes that $\bra\Gamma_1|J_{x,y}|\Gamma_4\ket=\bra\Gamma_2|J_{x,y}|\Gamma_3\ket=0$ so that we can 
restrict to $J_z$ in the model for inter-site interactions (Sec.~\ref{sect:mexcitons}).
If the singlet representations in the TSM would all be different then only one matrix element of $J_z$ could be non-zero. However, in the $J=4$ $D_{4h}$ decomposition given above each singlet representation occurs at least twice. Then two matrix elements of the TSM containing two singlets with equal symmetry can be non-zero. Because these have necessarily equal $\Theta$ the third $J_z$ matrix element is always zero as long as time reversal symmetry holds. In the induced magnetic phase when $\Theta$ is broken it will also be non-zero as shown in Appendix \ref{sect:app1} (Eq.~(\ref{eqn:matmag}); this is essential to obtain the proper temperature dependence of order parameter and soft mode energy. 

In the paramagnetic phase we are then
left with the three posible cases of dipolar matrix element sets  $(\alpha, \beta, \tal)$ as illustrated in Fig.~\ref{fig:TSM3}. Since the orthorhombic CEF is characterized by nine arbitrary CEF parameters one may reasonably expect that every sequence in Fig.~\ref{fig:TSM3} and similar ones with $(\Gamma_2,\Gamma_3)$ singlets can in principle be realized. We note that in the higher $D_{4h}$ symmetry only the model type of  Fig.~\ref{fig:TSM3}(c) seems possible (Appendix\ref{sect:app0}).
Rather than discussing each possible case presented in  Fig.~\ref{fig:TSM3} individually it is more economic to treat the general TSM (inset of  Fig.~\ref{fig:occ-level})  keeping in mind that always one in the set of matrix elements $(\alpha,\beta,\tal)$ must vanish to reproduce any of the possible cases in  Fig.~\ref{fig:TSM3} allowed by $\Theta$.

\section{Response function formalism, magnetic exciton bands and induced transtion}
\label{sect:excitons}

\subsection{Local dynamic susceptibility of the TSM}
\label{subsect:localsus}

The most direct way to understand the magnetic ordering in the TSM is provided by the response function formalism, the resulting magnetic exciton bands and their soft-mode behaviour. The dynamic response function for the isolated TSM is
in general given by
\bea
\chi_0(\om)=\sum_{ij}\frac{|\langle i|J_z|j \rangle |^2(p_j -p_i)}{E_i-E_j-\om}
\label{eqn:locchi}
\eea
defining the occupation differences of levels by $p_{ij}=p_i-p_j$ this is evaluated explicitly as $(\Delta_0=\Delta+\tDe)$
\bea
\chi_0(\om)=
2\Bigl[
\frac{\alpha^2\Delta p_{01}}{\Delta^2-(\om)^2} +
\frac{\tal^2\tDe p_{12}}{\tDe^2-(\om)^2} +
\frac{\beta^2\Delta_0 p_{02}}{\Delta_0^2-(\om)^2}
\Bigr]\nonumber\\
\label{eqn:locsus}
\eea
where the $p_{ij}$ are given by
\bea
p_{01}=\frac{\tanh\frac{\Delta}{2T}}{1-f_{01}};\;
p_{02}=\frac{\tanh\frac{\Delta_0}{2T}}{1+f_{02}};\;
p_{12}=\frac{\tanh\frac{\tDe}{2T}}{1+f_{12}}
\label{eqn:occdiff}
\eea
with
\bea
f_{01}&=&\fs(\cosh\frac{\tDe}{T}-\sinh\frac{\tDe}{T})(\tanh\frac{\Delta}{2T} -1)\non\\
f_{02}&=&\fs(\cosh\frac{\De}{T}-\sinh\frac{\De}{T})(\tanh\frac{\Delta_0}{2T} +1)\non\\
f_{12}&=&\fs(\cosh\frac{\Delta}{T}+\sinh\frac{\Delta}{T})(\tanh\frac{\tDe}{2T} +1)
\label{eqn:faux}
\eea
The occupation differences fulfil the relation $p_{12}=p_{02}-p_{01}$. For $T\ll \Delta,\Delta_0$
when $p_{01},p_{02}\leq 1$ this means $p_{12}\ll p_{01},p_{01}$.
For the {\it two}-singlet model $(i,j=0,1)$ one simply has $p_{ij}=\tanh\beh\Delta_{ij}$. 
In the TSM the expressions $f_{ij}$ in the denominators of Eq.(\ref{eqn:occdiff})
are a correction taking into account the presence of the third level in the partition function.

\subsection{Collective magnetic exciton modes}
\label{sect:mexcitons} 

The relevant part of the inter-site exchange interaction of three-singlet states is given by
 ($l,l'$ denote lattice sites $\bR_l,\bR'_l$):
\bea
H_{ex}&=&-\fs\sum_{ll'}I_{ll'}J_{z}(l)J_{z}(l')\nonumber\\
&=&-\fs\sum_\bq I_e(\bq)J_z^\bq J_z^{-\bq}
\label{eqn:hex}
\eea
with the Fourier component $J_z^\bq=N^{-\fs}\sum_l\exp(i\bq\bR_l)J_z^l$. The transverse  $J^\bq_{x}, J^\bq_{y}$ do not contribute to the collective mode dispersion because of
their vanishing matrix elements in the orthorhombic TSM's of Fig.~\ref{fig:TSM3}. The Fourier transform of the exchange interaction may be
expressed (assuming only next neighbor coupling $I_0$) as
\bea
I_e(\bq)=2I_0\gamma_\bq; \;\;\; \gamma_\bq=\sum_{n=1}^D\cos q_n
\label{eqn:iex}
\eea
in the simple orthorhombic lattice of dimension $D=3$ and coordination $z=2D$. The momentum units are $\frac{1}{a}$, $\frac{1}{b}$  and $\frac{1}{c}$ parallel to the respective orthogonal axes. For the AF case with $I_0<0$ on which we focus we also introduce
the effective AF exchange $I_e\equiv I_e(\bQ)=-zI_0 >0$ where $\bQ=(\pi,\pi,\pi)$ denotes the AF wave vector.
Then within RPA approximation \cite{jensen:91} the collective dynamic susecptiblity
(zz-component only) of coupled three singlet levels is obtained as
\bea
\chi(\bq,\om)=[1-I_e(\bq)\chi_0(\om)]^{-1}\chi_0(\om)
\label{eqn:RPAsus}
\eea
Its poles as defined by $1-I_e(\bq)\chi_0(\om)=0$ give the dispersive collective magnetic exciton modes of the  TSM which are determined by a cubic equation in $\omega^2$. We first derive  its general solution and then a more intuitive restricted one for the low temperature case.
%
% %%%%%%%%%%%%%%%%%%%%% figure %%%%%%%%%%%%%%%%%%%%%%%%%%%%
\begin{figure}
\vspace{1cm}
\includegraphics[width=0.90\columnwidth]{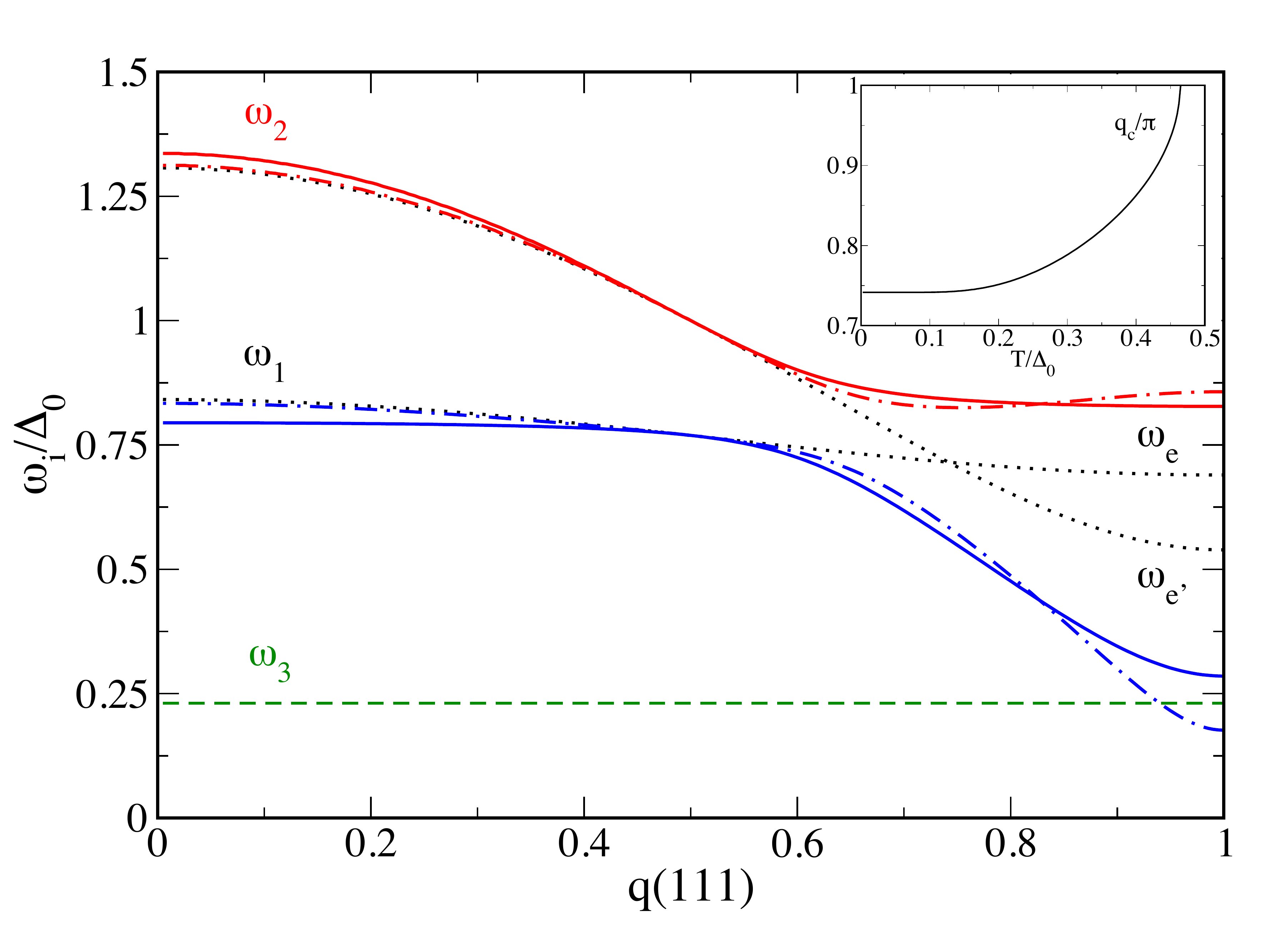}
\caption{Magnetic exciton dispersions along $(111)$ direction, $q_i$ is in units of $\pi/a$, $\pi/b$, $\pi/c$ and $T/\Delta_0=0.1$. Dotted black lines:  uncoupled modes where the upper one has stronger dispersion due to $\beta > \alpha$. Full blue/red lines are the coupled RPA modes (Eqs.~(\ref{eqn:modedisp},\ref{eqn:RPAmode})). The flat band (dashed green) corresponding to $1\leftrightarrow 2$ excitations has vanishing intensity (Fig.\ref{fig:LogSpec3}). The hybridization gap is determined by  the cross-coupling $\sim \alpha\beta$ (Eq.~(\ref{eqn:hybgap})). Dash-dotted blue/red lines show the dispersion for the main $\omega_{1,2}$ modes  in Bogoliubov approach (Eq.~(\ref{eqn:modedisp2})). Inset: crossing wave number $q_c$. Here $r=0.3$, $\xi_\alpha=0.2, \xi_\beta=0.71$   ($\xi_s=0.91<1$ subcritical), $\xi_{\tal}=0.$,  (scheme as Fig.~\ref{fig:TSM3}(a)).
}
\label{fig:RPA-bogol}
\end{figure}
%%%%%%%%%%%%%%%%%%%%%%fig%%%%%%%%%%%%%%%%%%%%%%%%%%%%%%%
%

\begin{figure}
\vspace{1cm}
\includegraphics[width=1.0\columnwidth]{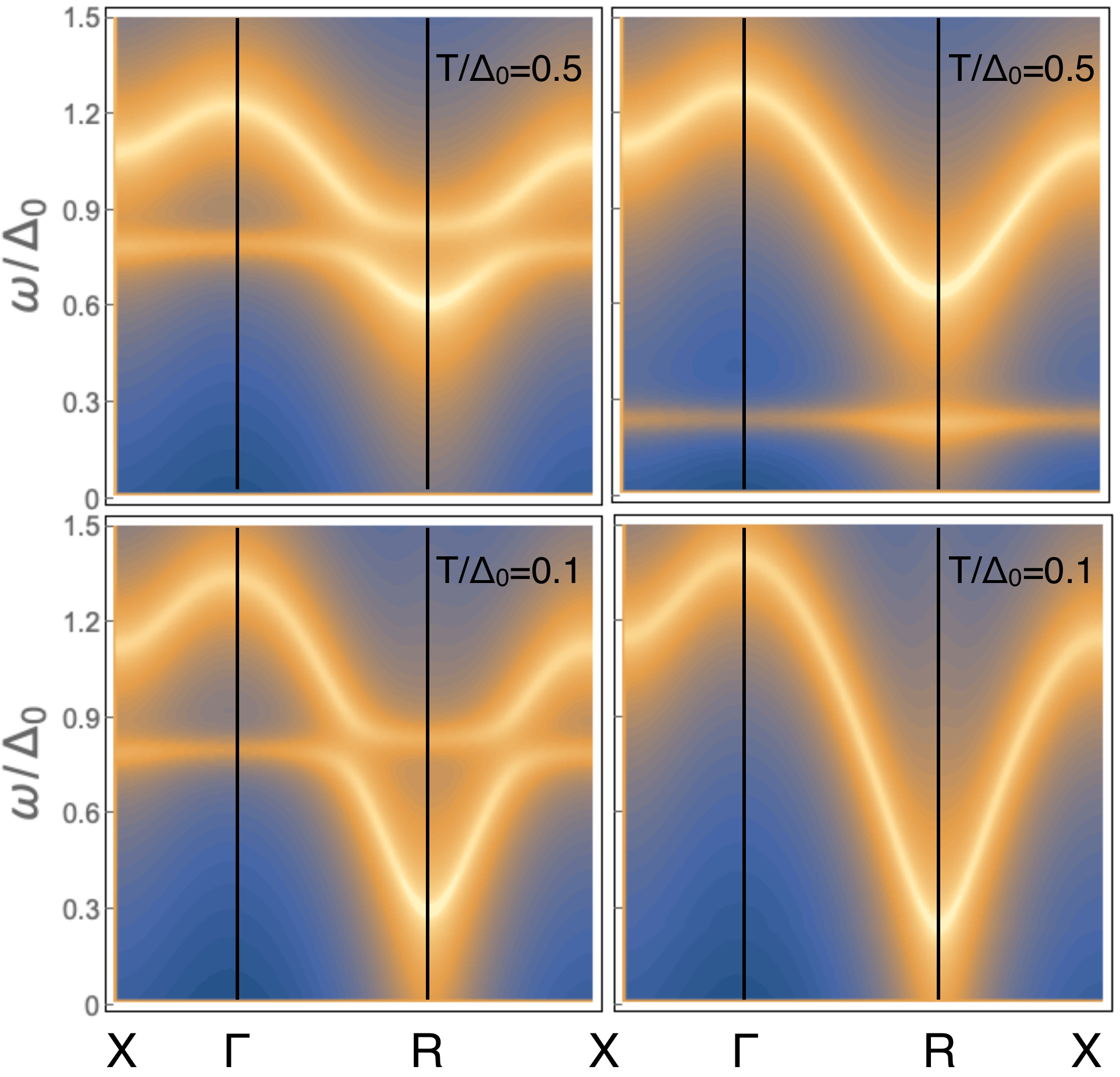}
\caption{Spectral density (log scale) of $\chi(\bq,\omega)$ along scaled orthorhombic BZ path X$(\pi,0,0)$, $\Gamma(0,0,0)$, R$(\pi,\pi,\pi)$, X$(\pi,0,0)$ for two temperatures, using the residua given in (Eqs.~(\ref{eqn:RPAspectral3},\ref{eqn:RPAspectral2})). Bright/dark colors correspond to large/small intensity. Left column: (scheme as Fig.~\ref{fig:TSM3}(a), parameters as Fig.~\ref{fig:RPA-bogol}) Hybridising modes $\omega_{1,2}$ with anti-crossing are shown.
For $T/\Delta_0=0.5$ dispersions are moderate, at $T/\Delta_0=0.1$ an incipient soft mode from $\omega_1(\bq)$ appears at the AF point (R) $\bQ=(\pi,\pi,\pi)$  (cf. Fig.~\ref{fig:RPA-bogol}). Right column: (scheme as  Fig.~\ref{fig:TSM3}(b), parameters 
 $r=0.3$, $\xi_\beta=0.94,\xi_{\tal}=0.66$, $\xi_{\alpha}=0$). For $T/\Delta_0=0.1$ a thermally excited flat mode $(\omega_3)$ appears. At $T/\Delta_0=0.1$ the latter is absent due to $p_{12}\ll1$ and dispersion of $\omega_{2}(\bq)$ shows
incipient soft mode behaviour. A spectral broadening of $\eta=0.025$ is used.}
\label{fig:LogSpec3}
\end{figure}

If we had only one of each contribution in Eq.~(\ref{eqn:locsus}) we would obtain
isolated exciton modes given by 
\bea
\omega^2_e(\bq,T)&=&\Delta[\Delta-2\alpha^2_TI_e(\bq)]\nonumber\\
\omega^2_{e'}(\bq,T)&=&\Delta_0[\Delta_0-2\beta^2_TI_e(\bq)]\nonumber\\
\omega^2_{e''}(\bq,T)&=&\tDe[\tDe-2\tal^2_TI_e(\bq)]
\label{eqn:baremode1}
\eea
where we used the effective T-dependent transition strengths defined by 
\bea
\alpha^2_T=\alpha^2p_{01},
\beta^2_T=\beta^2p_{02}, \tal^2_T=\alpha^2p_{12}. 
\label{eqn:Tcontrol}
\eea
These uncoupled modes are hybridized into new eigenmodes  when more matrix elements are present.
We derive these expressions already in sight of the magnetic case of  Sec.~\ref{subsect:AFexc} where all the
$\alpha'^2_T, \beta^2_T, \tal'^2_T$ as modified by the molcular field are nonzero.
The hybridized modes may be expressed in terms of the following auxiliary quantities:
\bea
\epsilon_1&=&-[\omega^2_e(\bq)+\omega^2_{e'}(\bq)+\omega^2_{e''}(\bq)]\nonumber\\
\epsilon_2&=&\omega^2_e(\Delta_0^2+\tDe^2)+\omega^2_{e'}(\Delta^2+\tDe^2)+\omega^2_{e''}(\Delta_0^2+\Delta^2)\nonumber\\
&&-(\Delta^2\Delta_0^2+\Delta^2\tDe^2+\Delta_0^2\tDe^2)\\
\epsilon_3&=&2\Delta^2\Delta_0^2\tDe^2-(\omega^2_e\Delta_0^2\tDe^2+\omega^2_{e'}\Delta^2\tDe^2+
\omega^2_{e''}\Delta^2\Delta_0^2)\nonumber
\eea
With the definition of 
\bea
P&=&\epsilon_2-\frac{1}{3}\epsilon_1^2;\;\;\;Q=\frac{2}{27}\epsilon_1^3-\frac{1}{3}\epsilon_1\epsilon_2+\epsilon_3\nonumber\\
\phi&=&\cos^{-1}\Bigl[-\frac{Q}{2}\bigl(\frac{|P|}{3}\bigr)^{-\frac{3}{2}}\Bigr]
\eea
The dispersions of of the coupled modes $(i=1,2,3)$ are given by
\bea
%\omega^2_1(\bq)&=&2\bigl(\frac{|P|}{3}\bigr)^{\frac{1}{2}}\cos\frac{\phi}{3}-\frac{\epsilon_1}{3}\nonumber\\
\omega^2_i(\bq)&=&2\bigl(\frac{|P|}{3}\bigr)^{\frac{1}{2}}\cos(\frac{\phi}{3}+\phi_i)-\frac{\epsilon_1}{3}
%\omega^2_3(\bq)&=&2\bigl(\frac{|P|}{3}\bigr)^{\frac{1}{2}}\cos(\frac{\phi}{3}+\frac{4\pi}{3})-\frac{\epsilon_1}{3}
\label{eqn:RPAmode}
\eea
where $\phi_1=0,\; \phi_2=\frac{2\pi}{3},\; \phi_3=\frac{4\pi}{3}$. 
These expressions give the RPA mode dispersions for any splittings and matrix elements of the TSM and also for abitrary temperature. In terms of these modes the collective RPA susceptibility may be written as
\bea
\chi(\bq,\omega)=
\frac{\chi_0(\omega)}{(\omega^2-\omega^2_1(\bq))(\omega^2-\omega^2_2(\bq))(\omega^2-\omega^2_3(\bq))}\nonumber\\
\label{eqn:dynsus}
\eea
This leads to a spectral function which determines the  structure function in inelastic neutron scattering (INS): 
\bea
\frac{1}{\pi}Im\chi(\bq,\omega+i\eta)=\sum_\lambda R_\lambda\delta(\omega-\omega_\lambda(\bq))
\eea
with the momentum and temperature dependent intensities $R_\lambda$ $(\lambda=1-3)$ of exciton modes given by
\bea
\label{eqn:RPAspectral3}
R_\lambda(\bq,T)&=&
\frac{Z(\omega_\lambda(\bq))}
{2\omega_\lambda(\bq)\Pi_{\mu\neq\lambda}[\omega^2_\mu(\bq)-\omega^2_\lambda(\bq)]}\\[0.2cm]
Z(\omega_\lambda(\bq))&=&
2\bigl[\alpha_T^2\Delta(\omega_\lambda(\bq)^2-\tDe^2)(\omega_\lambda(\bq)^2-\Delta_0^2)\nonumber\\
&&+\tal_T^2\tDe(\omega_\lambda(\bq)^2-\Delta^2)(\omega_\lambda(\bq)^2-\Delta_0^2)\nonumber\\
&&+\beta_T^2\Delta_0(\omega_\lambda(\bq)^2-\Delta^2)(\omega_\lambda(\bq)^2-\tDe^2)\bigr]\nonumber
%\label{eqn:RPAspectral3}
\eea
At low temperatures we can find an approximate and more intuitive solution for the dispersions:
For $T\ll \De,\De_0$ when $p_{12}\ll p_{01},p_{02}$ we can neglect the second term in Eq.~(\ref{eqn:locsus}), i.e. the influence of transitions starting from the thermally  excited states on the dynamics. Then the mode dispersions are  obtained in concise form as 
\bea
\omega_{1,2}^2(\bq)&=&\fs(\omega^2_e+\omega^{2}_{e'})\nonumber\\
&&\pm[\frac{1}{4}(\omega^2_e-\omega^{2}_{e'})^2+(2\alpha_T\beta_T)^2(I_e\Delta_{av})^2]^\fs
%\omega_e(\bq)&=&[\De(\De-2\alpha_T^2I_e(\bq))]^\fs\nonumber\\
%\omega_{e'}(\bq)&=&[\De_0(\De_0-2\beta_T^2I_e(\bq))]^\fs
\label{eqn:modedisp}
\eea
with $\Delta_{av}=(\Delta\Delta_0)^\fs$. The two dispersive modes stemming from the ground- to excited state transitions may anti-cross if their dispersion is sufficiently strong, i.e. if $I_0$ inter-site exchange is sufficiently large and matrix element $\alpha$ or $\beta$ large and sufficiently different. This happens  when the decoupled dispersions fulfil $\omega_e(\bq_c)=\omega_{e'}(\bq_c)\equiv \omega_c$. For $\bq=(q,q,q)$ along $\Gamma$R in the orthorhombic BZ where dispersion is maximal one obtains
\bea
q_c(T)&=&\cos^{-1}\Bigl(\frac{1}{12I_0}\frac{\Delta_0^2-\Delta^2}{\beta_T^2\Delta_0-\alpha_T^2\Delta}\Bigr)
%\;\; \text{if}\;\;12|I_0|>\Bigl |\frac{\Delta_0^2-\Delta^2}{\beta_T^2\Delta_0-\alpha_T^2\Delta}\Bigr | \nonumber\\
\eea
if the modulus of the argument is smaller than one. At the anti-crossing point $q_c$ of the two exciton modes (Figs.~\ref{fig:RPA-bogol},\ref{fig:LogSpec3}) the splitting otained from Eq.(\ref{eqn:modedisp})  is then given by
\bea
\delta\omega_c=\omega_1(q_c)-\omega_2(q_c)=
2|\alpha_T\beta_T| |I_e(q_c)|\frac{\Delta_{av}}{\omega_c}
\label{eqn:hybgap}
\eea
The anti-crossing happens because both inelastic transitions start from the same ground state and the splitting is therefore $\sim |\alpha_T\beta_T|$. The dispersion as well as the splitting decrease with increasing T due to the reduction of effective
transition strength $\alpha_T\sim p_{01}$ and $\beta_T\sim p_{02}$ (Fig.~\ref{fig:occ-level}). The intensities determining the spectral functions now take on the simplified form
\bea
\label{eqn:RPAspectral2}
R_\lambda(\bq,T)&=&
\frac{Z(\omega_\lambda)}
{2\omega_\lambda(\bq)[\omega^2_{\bar{\lambda}}(\bq)-\omega^2_\lambda(\bq)]}\\[0.2cm]
Z(\omega_\lambda)&=&2\bigl[\alpha_T^2\Delta(\omega_\lambda(\bq)^2-\Delta_0^2)+\beta_T^2\Delta_0(\omega_\lambda(\bq)^2-\Delta^2)\bigl]\nonumber
%\label{eqn:RPAspectral2}
\eea
where $\bar{\lambda}=2,1$ for $\lambda=1,2$, respectively. A discussion of exciton mode dispersions and intensities 
is given at the end of Sec.~\ref{sect:bogtrans}.
%
% %%%%%%%%%%%%%%%%%%%%% figure %%%%%%%%%%%%%%%%%%%%%%%%%%%%
%%%%%%%%%%%%%%%%%%%%%%fig%%%%%%%%%%%%%%%%%%%%%%%%%%%%%%%
%
\section{Diagonalization by Bogoliubov transformation and excitonic Bloch states}
\label{sect:bogtrans}

The response function formalism leads to a transparent picture for the excitonic mode dispersions, however 
it gives no information on the Bloch functions of these modes. For that purpose a direct (approximate) diagonalization
of the Hamiltonian using pseudo-unitary Bogoliubov and subsequent unitary transformations may be performed that
also contain the eigenvectors of exciton modes. Therefore we also apply this alternative approach to the problem. In this context the local CEF excitation standard basis operators $|i\rangle\langle j|$ in the TSM are mapped to bosons (altough the former have more complicated commutation relations). This  can be justified as long as the temperature fulfills $T\ll\Delta,\Delta_0$ and only the two excitations from the ground state  have to be considered \cite{grover:65,thalmeier:94} corresponding to the TSM of Fig.~\ref{fig:TSM3}(a). Defining $a^\dag_i=|1\rangle\langle 0|$  and $b^\dag_i=|2\rangle\langle 0|$ and using the Fourier transforms
$x^\dagger_\bk=(1/\sqrt{N})\sum_i\exp(-i\bk\bR_i)x^\dag_i$ $(x=a,b)$ the Hamiltonian $H=H_{CEF}+H_{ex}$ may be written in its bosonic form by using the definition $\psi^\dag_\bk =(a^\dag_\bk, a_{-\bk}, b^\dag_\bk, b_{-\bk})$ as
$H=\sum_\bk\psi^\dag_\bk h_\bk\Psi_\bk +\fs N(\Delta+\Delta_0)$ where (\bk~suppressed on right side.):
\be
\bl
&
h_\bk=\fs
\left(
 \begin{array}{cccc}
\Omega_a& -\alpha^2I_e& -\alpha\beta I_e& -\alpha\beta I_e\\
 -\alpha^2I_e& \Omega_a & -\alpha\beta I_e&  -\alpha\beta I_e \\
-\alpha\beta I_e&  -\alpha\beta I_e&\Omega_b&-\beta^2 I_e\\
-\alpha\beta I_e&  -\alpha\beta I_e&-\beta^2I_e&    \Omega_b
\end{array}
\right)
%\nonumber
\label{eqn:hambos}
\el
\ee
Here we defined 
\bea
\Omega^a_{\bk}=\Delta-\alpha^2I_e^\bk;\;\;\;\Omega^b_{\bk}=\Delta_0-\beta^2I_e^\bk.
\eea
This Hamiltionian may be approximately diagonalised by pseudounitary Bogoliubov transformations in each particle-hole subspace of $a,b$ - type operators  and a subsequent unitary rotation in the space of isolated A,B normal modes. The former are given  by
\bea
A_\bk&=&\cosh\theta^a_\bk a_\bk+\sinh\theta^a_\bk a^\dagger_{-\bk}\nonumber\\
B_\bk&=&\cosh\theta^b_\bk b_\bk+\sinh\theta^b_\bk b^\dagger_{-\bk}
\label{eqn:transbo}
\eea
which preserve the bosonic commutation relations for the $A_\bk, B_\bk$. The above transformation diagonalizes each diagonal $2\times 2$ block in Eq.~(\ref{eqn:hambos}) when the conditions
\bea
\tanh 2\theta^a_\bk&=&-\frac{\alpha^2I_e^\bk}{\Omega^a_\bk}=\frac{\Omega^a_\bk-\Delta}{\Omega^a_\bk}\nonumber\\
\tanh 2\theta^b_\bk&=&-\frac{\beta^2I_e^\bk}{\Omega^b_\bk}=\frac{\Omega^b_\bk-\Delta_0}{\Omega^b_\bk}
\label{eqn:theta}
\eea
are fulfilled. This leads to the transformed  Hamiltionian (in A,B particle space only) in terms of A,B uncoupled normal mode coordinates
given by
\bea
H&=&E_0+\sum_\bk(A^\dagger_\bk,B^\dagger_\bk)
\left(
 \begin{array}{cc}
\omega^A_\bk& 2\tl_\bk\\
 2\tl_\bk& \omega^B_\bk
\end{array}
\right)
\left(
 \begin{array}{c}
A_\bk\\
B_\bk
\end{array}
\right)\nonumber\\
E_0&=&\fs\sum_\bk[(\omega^A_\bk-\Delta)+(\omega^B_\bk-\Delta_0)]
\label{eqn:ABham}
\eea
Here $\omega^A_\bk $ and $\omega^B_\bk$ are the uncoupled normal mode frequencies 
\bea
\omega^A_\bk=[\Delta(\Delta-2\alpha^2I_e^\bk)]^\fs;
\omega^B_\bk=[\Delta_0(\Delta_0-2\beta^2I_e^\bk)]^\fs
\label{eqn:baremode2}
\eea
which are indeed equivalent to the uncoupled exciton modes $\omega_e, \omega_{e'}$, respectively of the RPA
response function approach in Eq.~(\ref{eqn:baremode1}) for the low temperature limit. 
Furthermore they satisfy the relations
\bea
\Omega^{a2}_\bk-\omega^{A2}_\bk=(\Omega^{a}_\bk-\Delta)^2;\;\;
\Omega^{b2}_\bk-\omega^{B2}_\bk=(\Omega^{b}_\bk-\Delta_0)^2 \nonumber\\
\eea
The coupling term in Eq.(\ref{eqn:ABham}) obtained through the transformation described by Eq.~(\ref{eqn:transbo}) is given by
\bea
\tl_\bk=-\fs(\alpha\beta)I^\bk_e(u^a_\bk-v^a_\bk)(u^b_\bk-u^b_\bk)
\eea
with $u^{a,b}_\bk=\cosh\theta^{a,b}_\bk$, $v^{a,b}_\bk=\sinh\theta^{a,b}_\bk$. It may be evaluated, 
using Eq.(\ref{eqn:theta}) as
\bea
\tl_\bk=-\fs\alpha\beta I_e^\bk \frac{\Delta^2_{av}}{\omega^A_\bk\omega^B_\bk}
\label{eqn:hybr}
\eea
Now a further unitary transformation in $A,B$ particle space can be employed according to
\bea
\chi_{1\bk}&=&\cos\phi_\bk A_\bk+ \sin\phi_\bk B_\bk\nonumber\\
\chi_{2\bk}&=&-\sin\phi_\bk A_\bk+ \cos\phi_\bk B_\bk
\label{eqn:transun}
\eea
%
%
% %%%%%%%%%%%%%%%%%%%%% figure %%%%%%%%%%%%%%%%%%%%%%%%%%%%
\begin{figure}
%\vspace{1cm}
\includegraphics[width=0.99\columnwidth]{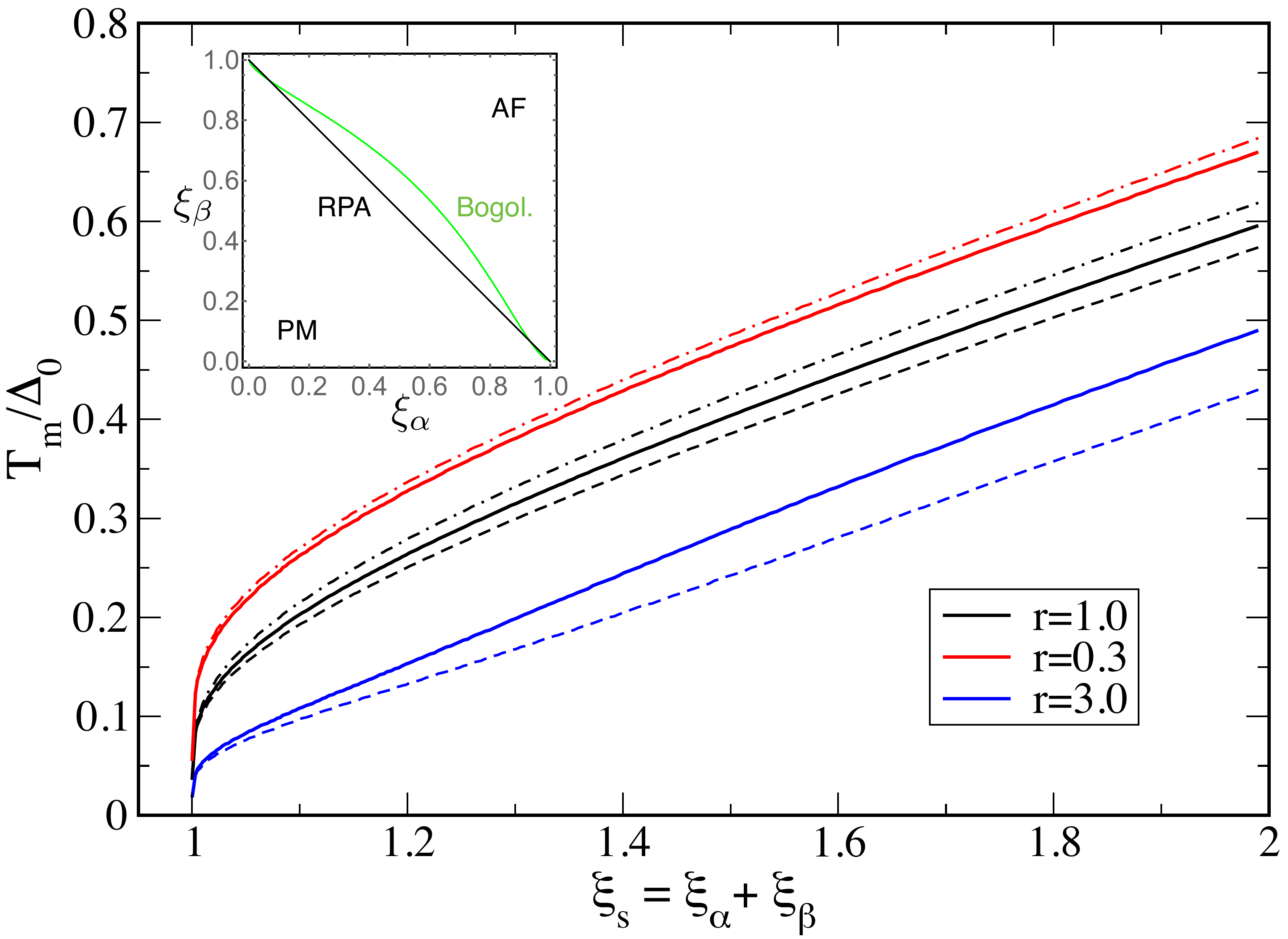}
\caption{The critical temperature for induced TSM order as function of control parameters $\xi_s$  for $\xi_a < 1$  and splitting
ratio $r=\tDe/\Delta$. Here full lines correspond to $\xi_a=0$ , dashed/dashed-dotted lines to $\xi_a=-0.2, 0.2$ (black) and 
$\xi_a=-0.3, 0.3$ (blue and red, respectively). Inset: PM/AF phase boundary $T_m(\xi_\alpha,\xi_\beta)=0$ for symmetric case $r=1$ from RPA (black) and Bogoliubov (green) theories.  }
\label{fig:tm2nd}
\end{figure}
%%%%%%%%%%%%%%%%%%%%%%fig%%%%%%%%%%%%%%%%%%%%%%%%%%%%%%%
%
These are the normal mode exciton coordinates that diagonalise the Hamiltonian in Eq.~(\ref{eqn:hambos})
(up to residual two-exciton interactions)
% $(H_{(2)}$ in Eq.~(\ref{eqn:hamfin}), see Appendix \ref{sect:app2}) 
provided the condition
\bea
\tan 2\phi_\bk=\frac{4\tl_\bk}{\omega^A_\bk-\omega^B_\bk}=
\pm\Bigl[\bigl(\frac{\omega_{1\bk}-\omega_{2\bk}}{\omega^A_\bk-\omega^B_\bk}\bigr)^2-1\bigr]^\fs
\label{eqn:phi}
\eea
is fulfilled, leading to 
\bea
H=E_0+\sum_\bk[\omega_{1\bk}\chi^\dagger_{1\bk}\chi_{1\bk}+\omega_{2\bk}\chi^\dagger_{2\bk}\chi_{2\bk}] 
%+H_{(2)}
\label{eqn:hamfin}
\eea
where the exciton mode frequencies are finally given by
\bea
\omega_{(1,2)\bk}=\fs(\omega^A_\bk+\omega^B_\bk)
\pm[\frac{1}{4}(\omega^A_\bk-\omega^B_\bk)^2+4\tl_\bk^2]^\fs
\label{eqn:modedisp2}
\eea
which essentially corresponds to the RPA result of Eq.~(\ref{eqn:modedisp}) for zero temperature.
Obviously the direct
diagonalization route to obtain the exciton modes is more elaborative than the response function
formalism. On the other hand it also provides the Bloch functions $\chi^\dagger_{1,2\bk}|0\rangle$ 
whose creation operators are, according to Eqs.~(\ref{eqn:transbo},\ref{eqn:transun}) explicitly given by
\bea
\chi^\dagger_{1\bk}&=&c\tch_aa^\dag_\bk+c\tsh_aa_{-\bk}+s\tch_bb^\dag_\bk+s\tsh_bb_{-\bk}\nonumber\\
\chi^\dagger_{2\bk}&=&-s\tch_aa^\dag_\bk-s\tsh_aa_{-\bk}+c\tch_bb^\dag_\bk+c\tsh_bb_{-\bk}
\label{eqn:normalcoord}
\eea

where we defined $s=\sin\phi_\bk, c=\cos\phi_\bk$ and $\tsh_{a,b}=\sinh\theta^{a,b}_\bk, \tch_{a,b}=\cosh\theta^{a,b}_\bk$.
They fulfil the standard bosonic commutation relations $[\chi_{n\bk},\chi^\dag_{n'\bk'}]=\delta_{nn'}\delta_{\bk\bk'}$ $(n=1,2)$.\\
%
% %%%%%%%%%%%%%%%%%%%%% figure %%%%%%%%%%%%%%%%%%%%%%%%%%%%
\begin{figure}
\includegraphics[width=0.99\columnwidth]{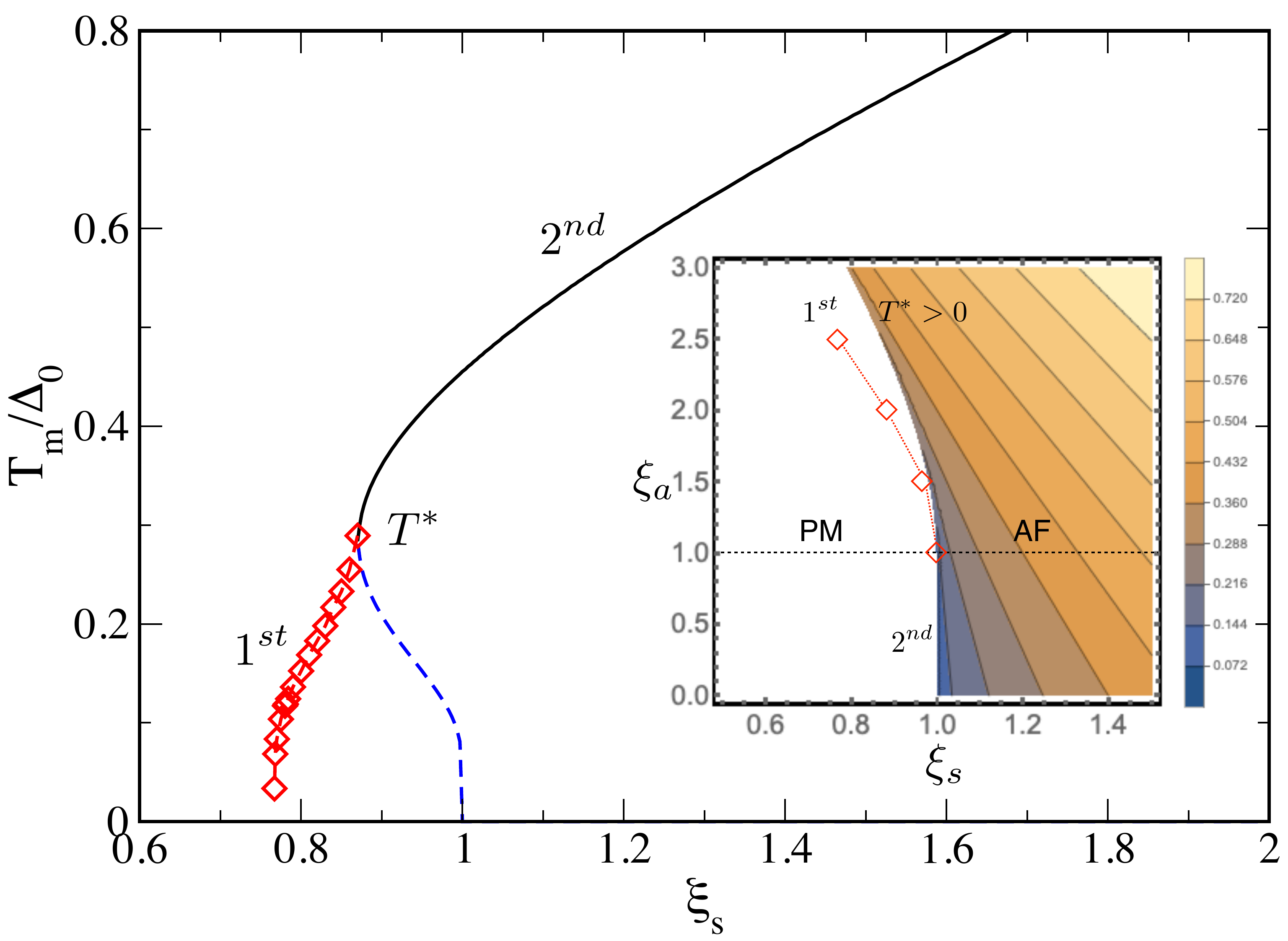}
\caption{
Critical temperature as function of $\xi_s$  for $\xi_a=2.5>1$  for the weakly symmetric case ($r=1$). Here full black line corresponds to $2^{nd}$ order transition obtained from physical solution of Eq.~(\ref{eqn:ym}) as in Figs.~\ref{fig:tm2nd},\ref{fig:OP-sus} and red symbols and dashed line denote a first order transition (cf.  Fig.~\ref{fig:OP-sus1}). Here $T^*$ denotes the critical point. The dashed blue line gives the lower unphysical solution of  Eq.~(\ref{eqn:ym}).
Inset: phase diagram of induced AF order (weakly symmetric $r=1$ TSM) in the control parameter
plane $(\xi_s,\xi_a)$. For $\xi_a<1$ the $T_m=0$ phase boundary does not depend on $\xi_a$ (cf. Fig.~\ref{fig:tm2nd} and Sec. \ref{sect:tm}).  For  $\xi_a>1$  $T^*$ maps the line of critical points between $1^{st}/2^{nd}$ order regime and the red symbols trace the first order transition line $T_m=0$.
The blue lower corner corresponds to the inverse logarithmic decrease of  $T_m/\Delta_0$ evident from Fig.~\ref{fig:tm2nd}.}
\label{fig:tm1st-inset}
\end{figure}
%%%%%%%%%%%%%%%%%%%%%%fig%%%%%%%%%%%%%%%%%%%%%%%%%%%%%%%
%
We can give explicit expressions for the transformation coefficients in Eq.~(\ref{eqn:normalcoord}) in terms of the various isolated and coupled eigenmode frequencies  by eliminating the 
angles $\theta^{a,b}_\bk$ and $\phi_\bk$. Using Eqs.(\ref{eqn:theta},\ref{eqn:phi}) we obtain
\bea
\tch_{a,b}=\bigl[\fs(\frac{\Omega^{a,b}_\bk}{\omega^{A,B}_\bk}+1)\Bigr]^\fs;\;\;
\tsh_{a,b}=\bigl[\fs(\frac{\Omega^{a,b}_\bk}{\omega^{A,B}_\bk}-1)\Bigr]^\fs
\eea
for the Bogoliubov transformation coefficients. Likewise we get
\bea
c=\bigl[\fs(1+\frac{|\omega^A_\bk-\omega^B_\bk|}{|\omega_{1\bk}-\omega_{2\bk}|})\bigr]^\fs;\;\;
s=\bigl[\fs(1-\frac{|\omega^A_\bk-\omega^B_\bk|}{|\omega_{1\bk}-\omega_{2\bk}|})\bigr]^\fs \nonumber\\
\eea
for the coefficients of the subsequent unitary transformation.

The comparison of the low  excitonic modes at low temperature as obtained from response function  and Bogoliubov approach is shown
in Fig.~\ref{fig:RPA-bogol} for the $(111)$ direction. Control parameters are chosen such that a crossing of uncoupled modes (dotted lines) of Eqs.~(\ref{eqn:baremode1},\ref{eqn:baremode2}) occurs at wave number $q_c$. Their hybridisation leads to an anticrossing of
the coupled modes (Eqs.~(\ref{eqn:RPAmode},\ref{eqn:modedisp},\ref{eqn:modedisp2})). The full line  represents
RPA result of Eqs.~(\ref{eqn:RPAmode},\ref{eqn:modedisp}). The inset 
depicts the increase of the crossing wave number with temperature. Once it has reached the zone boundary the modes become
gradually decoupled due the suppression of their dispersion.
At the AF zone boundary vector \bQ~$(q/\pi=1)$ the lower mode $\omega_1(\bq)$ shows incipient softening. The dash-dotted line is obtained from the Bogoliubov result in Eq.~(\ref{eqn:modedisp2}) and is rather close to the full line. There are, however distinct
differences close to the AF point: Because the effective hybridisation $\tl_\bk$ (Eq.~(\ref{eqn:hybr})) is enhanced by a feedback effect due to the mode softening, the latter happens more rapidly in the Bogoliubov approach. This will also lead to a difference in
the phase boundary for the two techniques (inset of Fig.~\ref{fig:tm2nd}). 
%
% %%%%%%%%%%%%%%%%%%%%% figure %%%%%%%%%%%%%%%%%%%%%%%%%%%%
\begin{figure}
\vspace{1cm}
\includegraphics[width=0.99\columnwidth]{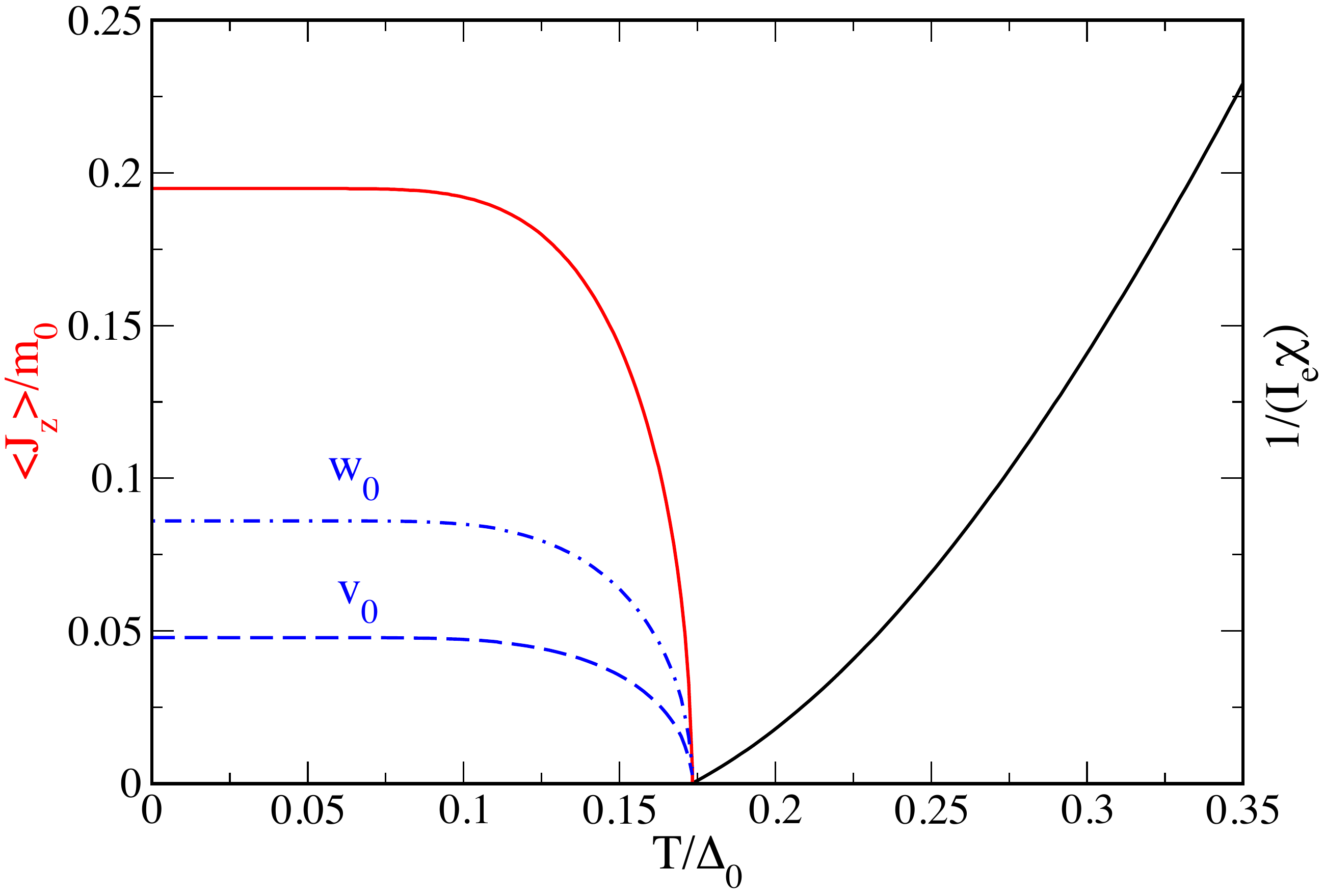}
\caption{Temperature dependence of inverse normalized susceptibility  in the paramagnetic phase (black) and
order parameter (induced moment $\langle J_z\rangle$) in the AF phase (red) normalized to $m_0=(\alpha^2+\beta^2)^\fs$. 
The transition for $\xi_s =1.02$ and $\xi_a= -0.2$ is of second order (Fig.~\ref{fig:tm2nd}).
The moment appears due to the mixing of excited $|1\rangle$ and $|2\rangle$ into the mf ground state $|\psi_0\rangle$ 
(Eq.~(\ref{eqn:mfstate})). Their coefficients $v_0,w_0$ are shown as broken blue lines (scheme as Fig.~\ref{fig:TSM3}(a)). }
\label{fig:OP-sus}
\end{figure}
%%%%%%%%%%%%%%%%%%%%%%fig%%%%%%%%%%%%%%%%%%%%%%%%%%%%%%%
%
The temperature dependence of spectral functions for TSM's of Fig.~\ref{fig:TSM3} (a),(b) obtained from RPA theory (Eq.~(\ref{eqn:RPAspectral3})) is presented in  Fig.~\ref{fig:LogSpec3}  (left and right columns, respectively) and shows distinctive features. Left: (i) With increasing temperature the anti-crossing region moves to larger wave vectors, concommitant with $q_c(T)$ in Fig.~\ref{fig:RPA-bogol}  (ii) With decreasing temperature $\omega_1(\bq)$ becomes an incipient soft mode. For slightly larger $\xi_\beta$ it would become unstable at lowest temperature. Correspondingly the dispersive width of $\omega_2(\bq)$ increases for lower temperature. Right: (i) At larger temperature the mostly flat $\omega_3(\bq)$ low-energy mode originating from transitions between thermally excited $|1\rangle$ and $|2\rangle$ states is still visible, its flatness is caused by the always small thermal population difference factor $p_{12}$ (Fig.~\ref{fig:occ-level}). For this reason its spectral weight also decays exponentially at low temperature and therefore it has vanished from Fig.~\ref{fig:LogSpec3}  $(T/\Delta_0=0.1$). (ii) The $\omega_2(\bq)$ mode now shows incipient soft mode behaviour due to slightly below-critical control parameters.

\section{Soft mode behaviour and critical condition for magnetic order}
\label{sect:tm}

When temperature is lowered the effective coupling parameters  $\alpha^2_T=\alpha^2p_{01}$,
$\beta^2_T=\beta^2p_{02}$ for the TSM of Fig.~\ref{fig:TSM3}(a) increase and with it the dispersive width of $\omega_1,\omega_2$ modes.
Eventually one of them may touch zero a the wave vector $\bq$ where $I_e(\bq)$ has its maximum,
frequently (but not necessarily) at zone center $\bq=0$ or boundary $\bQ=(\pi,\pi,\pi)$. This mode softening signifies the onset
of induced excitonic FM, AF quantum magnetism at $T_m$, respectively. In distinction to common magnetic order the moments are not
preformed already at larger temperature and order at $T_m$, since there are only nonmagnetic singlet states available,  
but rather the creation and ordering of moments happens simultaneously at $T_m$ due to off-diagonal virtual 
transitions between the singlets.\\
%
% %%%%%%%%%%%%%%%%%%%%% figure %%%%%%%%%%%%%%%%%%%%%%%%%%%%%%%%%%
\begin{figure}
\vspace{1cm}
\includegraphics[width=0.99\columnwidth]{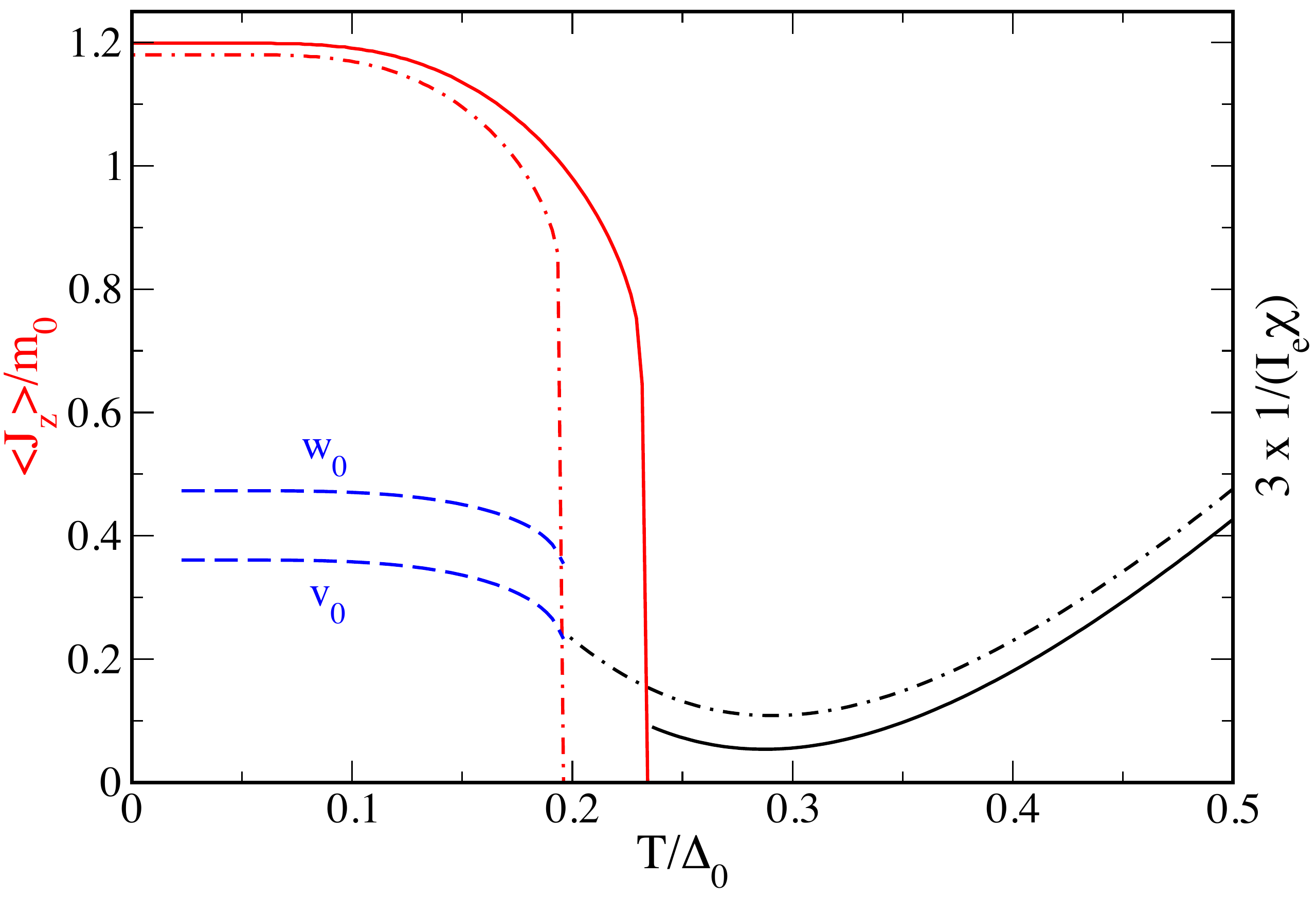}
\caption{Same as Fig.~\ref{fig:OP-sus}, now for  $\xi_s =0.85 , 0.83 $ ( full and broken black/red lines ) and $\xi_a= 2.5$ when the transition is of first order (Fig.~\ref{fig:tm1st-inset}) with jump in $\langle J_z\rangle$ . Note that susceptibility above $T_m$ does not diverge  due to first order character. Ground state admixture coefficients $v_0,w_0$ (broken blue lines) also jump to finite values at $T_m$  (scheme as Fig.~\ref{fig:TSM3}(b)). }
\label{fig:OP-sus1}
\end{figure}
%%%%%%%%%%%%%%%%%%%%%%fig%%%%%%%%%%%%%%%%%%%%%%%%%%%%%%%%%%%%
%
We first consider the soft mode condition within the RPA response function formalism. According to Eq.~(\ref{eqn:dynsus}) it is equivalent to the divergence of the static susceptibility $\chi(\bq,T_m)^{-1}\rightarrow 0$ which leads to the criticality condition 
\bea
\chi_0(0,T_m)=\frac{1}{I_e(\bq)}
\label{eqn:critical}
\eea
This means that the static $(\om=0)$ single-ion susceptibility given by Eq.~(\ref{eqn:locsus}) must reach a mininum value $\geq 1/I_e(\bq)$ to achieve induced magnetic order at finite $T_m$. We focus at the AF case $(I_0<0)$ where this is first fulfilled for the AF wave vector $\bq= \bQ$. The procedure for FM $(\bq=0)$ or even incommensurate cases are analogous.\\

As a reference we recapitulate the well known expression for $T_m$  in the {\it two-singlet} model \cite{wang:68,cooper:72,jensen:91,thalmeier:02}
 (e.g. taking off the upper singlet-state $|2\rangle$ in Fig.~\ref{fig:occ-level}). In this case $(I_e\equiv I_e(\bQ)=-zI_0>0)$:
 \bea
 T_m=\frac{\Delta}{2\tanh^{-1}\frac{1}{\xi_\alpha}}; \;\;\; \xi_\alpha=\frac{2\alpha^2I_e}{\Delta}
 \label{eqn:tm2}
 \eea
where $\xi_\alpha$ is now the only dimensionless control parameter of the model and at $\xi^c_\alpha=1$ a quantum phase transition from paramagnetic $\xi<\xi^c_\alpha$ to magnetic $\xi>\xi^c_\alpha$ ground state appears. In the marginally critical case $\xi_\alpha=1+\delta$ ($0<\delta\ll 1$) we can expand 
$T_m\simeq\Delta/\ln\bigl(\frac{2}{\delta}\bigr)$
and thus the ordering temperature vanishes logarithmically when approaching the critical value $\xi^c_\beta$ $(\delta\rightarrow 0)$ from above. This is a characteristic behaviour of an induced excitonic quantum magnet.\\
%
% %%%%%%%%%%%%%%%%%%%%% figure %%%%%%%%%%%%%%%%%%%%%%%%%%%%
\begin{figure}
\vspace{1cm}
\includegraphics[width=0.87\columnwidth]{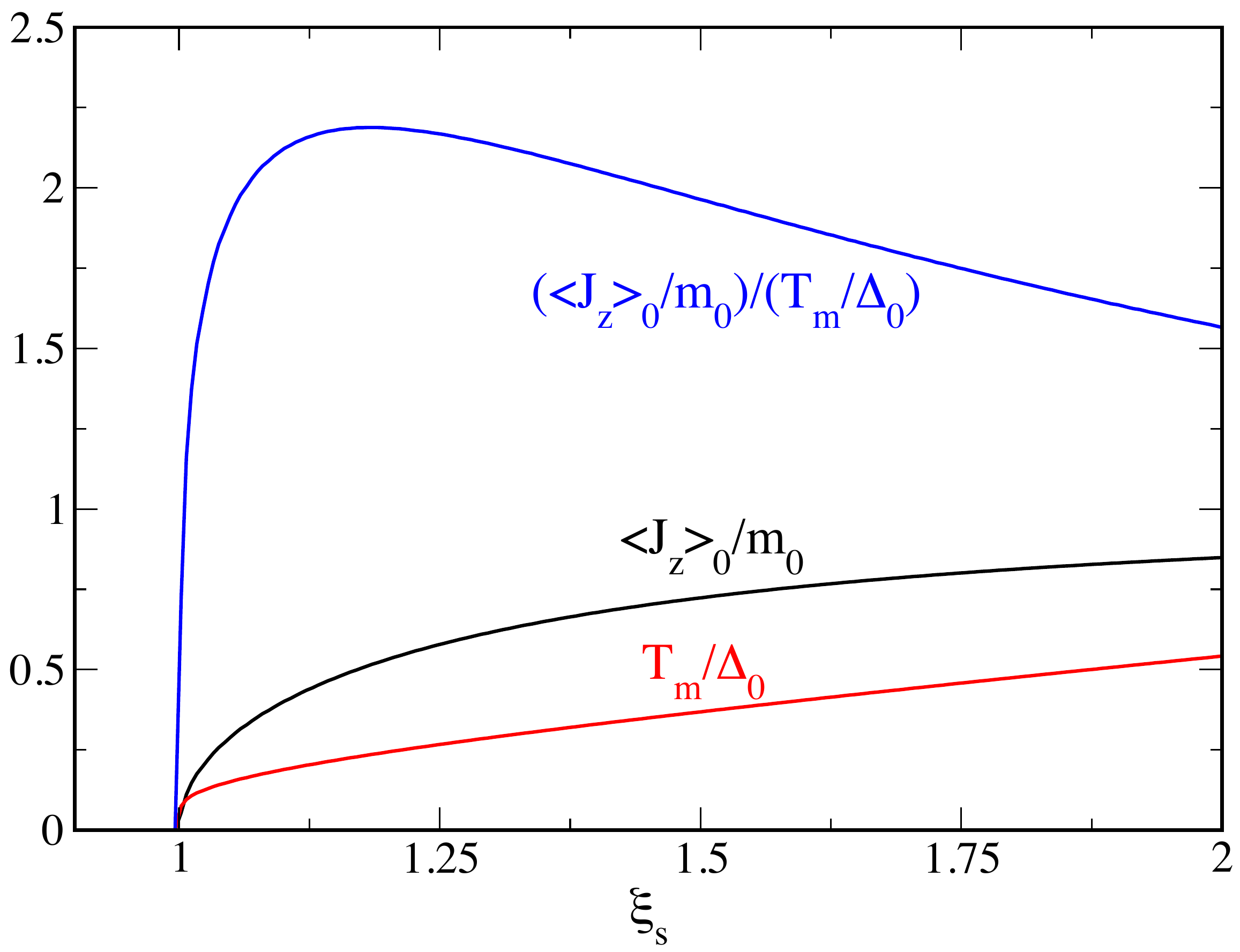}
\caption{Normalized saturation moment $\langle J_z\rangle_{T=0}/m_0$  with $m_0=(\alpha^2+\beta^2)^\fs$, transition temperature $T_m/\Delta_0$ and their ratio as function of $\xi_s$. Matrix elements $\alpha,\beta,\tal$ as in Fig.~\ref{fig:OP-sus} for weakly symmetric case with $r=1$. Here the interaction constant I$_e$ is varied leading to concommitant variation of $\xi_s$ and $\xi_a$.
The ratio shows steep decrease close to quantum critical point  (scheme as Fig.~\ref{fig:TSM3}(a)).}
\label{fig:satmom}
\end{figure}
%%%%%%%%%%%%%%%%%%%%%%fig%%%%%%%%%%%%%%%%%%%%%%%%%%%%%%%
%

Now we consider the extended TSM cases of  Fig.~\ref{fig:TSM3}  with generally possible  parameter sets.
The critical equation for $T_m$ (Eq.~(\ref{eqn:tm2})) may be written with the use of control parameters of Eq.~(\ref{eqn:control3}) as
\bea
\xi_\alpha p_{01}(T_m)+\xi_\beta p_{02}(T_m)+\xi_{\tal} p_{12}(T_m)=1
\eea
For convenience we now define the
splitting ratio $r=\tDe/\Delta$, meaning $\Delta=\Delta_0/(1+r)$. Then $r=1$ corresponds to the symmetric case with $\Delta=\tDe$ (Fig.~\ref{fig:occ-level}) and $r\neq1$ to the general asymmetric case. Defining furthermore $y=\exp(\Delta/T)$ the critical condition for induced order Eq.~(\ref{eqn:critical}) can be written as
\bea
(1-\xi_s)y^{r+1}_m+(1-\xi_a)y^r_m+(1+\xi_s+\xi_a)=0
\label{eqn:ym}
\eea
then $T_m=\Delta/\ln y_m$ is the ordering temperature with $y_m$ given by the solution of
the above algebraic equation. Note that even in the general asymmetric TSM  described by Eq.~(\ref{eqn:ym}) there are effectively
two control parameters which are combinations of the three possible parameters in Eq.~(\ref{eqn:control3}) according to $\xi_s=\xi_\alpha+\xi_\beta$ and $\xi_a=\xi_{\tal}-\xi_\alpha$ leading explicitly to the expressions
\bea
\xi_s=\frac{2I_e}{\Delta}(\frac{\beta^2}{1+r}+\alpha^2);\;\;\xi_a=\frac{2I_e}{\Delta}(\frac{\tal^2}{r}-\alpha^2)
\label{eqn:xisa}
\eea
We should remember that in the paramagnetic state one of the elements in the se $(\alpha,\beta,\tal)$  must be zero corresponding to the cases of Fig.~\ref{fig:TSM3}.
The solution of Eq.~(\ref{eqn:ym}) for finite $T_m$ and general splitting ratio $r$ is only possible numerically. However, for discrete values like $r=\fs, 1,2,3$ explicit solutions for $T_m$  can be obtained but except for $r=1$ are not particularly instructive. We may also look at the limiting cases $r\rightarrow 0,\infty$. The latter corresponds to the singlet-singlet model and recovers the solution in Eq.~(\ref{eqn:tm2}) while the former describes the singlet-doublet model \cite{bak:75} with splitting $\Delta$. Its $T_m$ is also described by  Eq.~(\ref{eqn:tm2}) but with the replacement $\xi_\beta\rightarrow 2(\alpha^2+\beta^2)I_e/\Delta$.\\

Now we discuss two typical special cases of the TSM model where the solution for $T_m$ can be obtained in closed form from Eq.~(\ref{eqn:ym}). These are considerably more complicated to derive than for the two level system but formally similar:\\

i) {\it weakly symmetric TSM}  $r=1$ but $\tal\neq \alpha$\\
Then Eq.~(\ref{eqn:ym}) reduces to a quadratic equation and from its two solutions $y_m^\pm$ the critical temperature may 
be obtained as
\bea
T^\pm_m&=&\frac{\Delta}{2\tanh^{-1}\frac{1}{\eta^\pm_s}};\;\;\; \frac{1}{\eta^\pm_s}=\frac{y_m^\pm-1}{y_m^\pm+1} 
\eea
Explicitly one obtains after some derivations:
\bea
\frac{1}{\eta^\pm_s}&=&\frac{2\xi_s+\xi_a\pm\bigl[(4\xi_s^2-3)+2\xi_a(2\xi_s-3)+\xi_a^2\bigr]^\fs}
{1+2\xi_a}
\nonumber
\label{eqn:etaex}
\eea
where $\xi_s,\xi_a$ are given by Eq.~(\ref{eqn:xisa}) with $r=1$. Instead of having directly the control parameter $\xi_s$ appearing in $T_m$ as in Eq.(\ref{eqn:tm2}) it is replaced by a function $\eta_s(\xi_s,\xi_a)$.  A solution for finite $T^\pm_m$ exists only when $\eta^\pm_s(\xi_s,\xi_a)\geq 1$. The physical solution is always $T_m\equiv T_m^-$ with $\eta_s=\eta_s^-$. The second solution $T_m^+$ does not exist for $\xi_a<1$ and for $\xi_a>1$ corresponds to the unphysical branch with $T_m^-<T_m$ (blue dashed line in Fig.~\ref{fig:tm1st-inset}). 
It is easy to show that $\eta_s^{-1}(1,\xi_a)=1$ for $\xi_a\leq 1$. Therefore the $T_m$ and the phase boundary position $\xi_s$ does not depend on $\xi_a$ in this case as is indeed demonstrated by Fig.~\ref{fig:tm2nd} and inset of Fig.~\ref{fig:tm1st-inset}.\\
%
% %%%%%%%%%%%%%%%%%%%%% figure %%%%%%%%%%%%%%%%%%%%%%%%%%%%
\begin{figure}
\vspace{1cm}
\includegraphics[width=0.95\columnwidth]{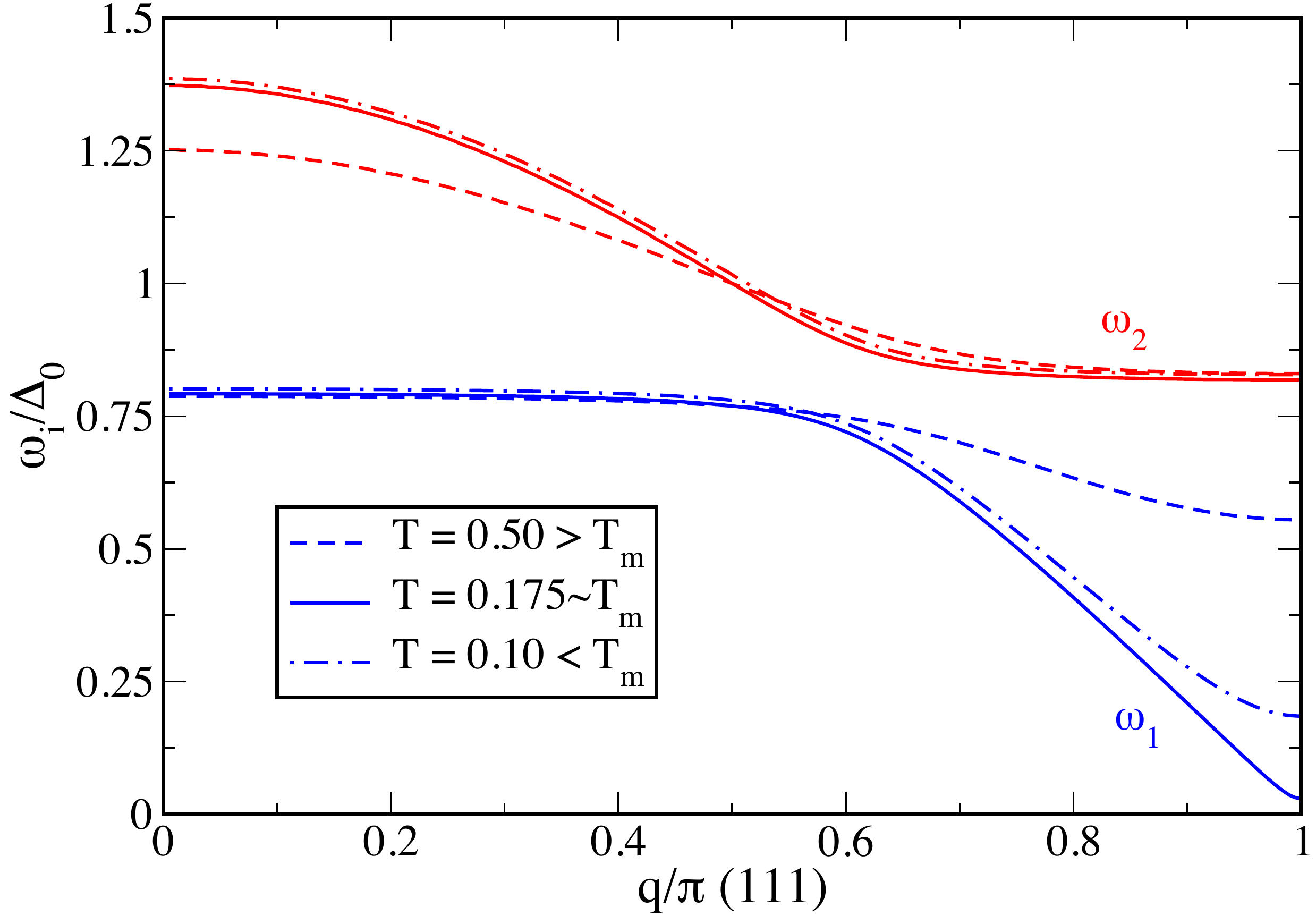}
\caption{Exciton dispersions for three temperature above, equal and below the AF 
transition temperature at $T_m\simeq 0.175$. The softening of the critical mode $\omega_2(\bq)$ at the zone
boundary AF point $\bQ=(\pi,\pi,\pi)$ is reversed below $T_m$ into a hardening with decreasing temperature 
(cf. Fig.~\ref{fig:softmode}). Here we have above-critical values $\xi_s=1.02, \xi_a=-0.2$  (scheme as Fig.~\ref{fig:TSM3}(a)).}
\label{fig:disptemp}
\end{figure}
%%%%%%%%%%%%%%%%%%%%%%fig%%%%%%%%%%%%%%%%%%%%%%%%%%%%%%%
%
ii) {\it fully symmetric TSM}  $r=1$ and $\tal = \alpha$\\
This means that now $\xi_a=0$ and only {\it one} effective control parameter $\xi_s$ remains. $T_m$ is given by the same expression as above but with the simplification
\bea
\eta^{-1}_s&=&[2\xi_s-[4\xi_s^2-3]^\fs]\nonumber\\             
\xi_s&=&\frac{2(\alpha^2+\fs\beta^2)I_e}{\Delta}
\label{eqn:tmsymm}
\eea
where now $\xi_s$ is the control parameter for the fully symmetric TSM that contains both matrix elements and the splitting $\Delta=\Delta_0/2$.   For a finite $T_m$ one must have $\eta_s>1$ and hence $\xi_s >1$. In the marginal critical case $\xi_s=1+\delta$ the transition temperature shows similar logarithmic behaviour as before, but with $T_m\simeq\Delta/\ln\bigl(\frac{1}{\delta}\bigr)$.

The systematic variation of $T_m(\xi_s,\xi_a;r)$  is shown in Fig.~\ref{fig:tm2nd} for $\xi_a<1$.  The fully and weakly symmetric cases $(r=1)$ discussed in detail above  correspond to the full and broken black lines in  Fig.~\ref{fig:tm2nd}, respectively. In the asymmetric case $(r\neq 1$) the transition temperature $T_m$ changes considerably with the splitting asymmetry $r=\tDe/\Delta$, keeping the total splitting $\Delta_0$ constant. When $r <1$ the central state $|1\rangle$ is shifted upwards leading to an increased effectiveness because the occupation difference $p_{01}$ increases, therefore $T_m$ increases. The reversed argument holds for $r>1$. Furthermore when the asymmetric control parameter $\xi_a$ is larger or smaller than zero for a given $r$ the value of $T_m$ moderately increases or decreases, respectively.  

For $\xi_a >1$ when the coupling of thermally excited states becomes important a surprising new situation occurs (Fig.~\ref{fig:tm1st-inset}): Firstly the second order transition temperature $T_m$ now stays finite for $\xi_s<1$ and secondly at a certain critical point $T_m=T^*$ it changes into a first order transition for $T_m<T^*$. This is of course no longer described by Eq.~(\ref{eqn:ym}) and its special cases since it was obtained from the divergence of the susceptibility at $T_m$. Below $T^*$ this is no longer true and $T_m$ has to be determined by solving directly the selfconsistency equations for the order parameter (Sec.~\ref{sect:AF}). The resulting line of first order transitions is shown by red symbols and dashed line in the main Fig.~\ref{fig:tm1st-inset} for $\xi_a=2.5$. For this value the $1^{st}$ order line stops at $\xi_s=0.77$.

Alternatively this variation can be combined in a contour plot of $T_m$ in the $(\xi_s,\xi_a)$ 
control parameter plane for fixed $r$, taken as the symmetric case $r=1$ in the inset of Fig.~\ref{fig:tm1st-inset}. 
Firstly it shows that the sector of first order transitions bounded by the red symbols and broken line to the left and
the $T^*>0$ line to the right widens when $\xi_a$ increases, i.e. the transitions between thermally excited states
become more important. Secondly it shows explicitly the $\xi_s$-independence of the second order PM/AF phase boundary defined by  $T_m(\xi_s,\xi_a;r)=0$ for $\xi_a<1$ as already noticed before. This property may be traced back directly to the fundamental equation for $T_m$ given by Eq.~(\ref{eqn:ym}). 

In this  respect it is instructive to compare 
the predictions of the soft mode conditons $\omega_2(\bQ)=0$ at $T_m=0$  for RPA (Eq.~(\ref{eqn:modedisp})) and Bogoliubov (Eq.~(\ref{eqn:modedisp2})) approaches for consistency (in the case $\tal=0$ of Fig.~\ref{fig:TSM3}(a)). They cannot be identical due to the slightly different expressions for the exciton mode dispersions. In the RPA case one simply obtains from  the equivalent Eq.~(\ref{eqn:ym}) in the limit $y_m\rightarrow \infty$: $\xi_s=\xi_\alpha+\xi_\beta=1$, in accordance with previous discussion of symmetric models (inset of Fig.~\ref{fig:tm1st-inset} for $\xi_a<1$). This means the effect of the two excitations is simply additive at the phase boundary. In comparison the Bogoliubov case leads to the more complicated  relation 
\bea
\bigl[(\xi_\alpha-1)(\xi_\beta-1)\bigr]=(\frac{1}{4}\xi_\alpha\xi_\beta)^\frac{1}{3}
\eea
For the special case $\xi_\alpha=\xi_\beta=\xi$ we obtain $\xi=0.5$ in the RPA approach and $\xi=0.57$ in the Bogoliubov approach. Furthermore in both cases the boundary points $(\xi_\alpha,\xi_\beta)= (1,0), (0,1)$ are identical for both methods. The complete comparison of PM/AF phase boundaries $T_m(\xi_\alpha,\xi_\beta)=0$ is shown in the inset of   Fig.~\ref{fig:tm2nd} for both methods. It demonstrates a rather close agreement between the two technically rather different approaches.\\
%
% %%%%%%%%%%%%%%%%%%%%% figure %%%%%%%%%%%%%%%%%%%%%%%%%%%%
\begin{figure}
\vspace{1cm}
\includegraphics[width=0.95\columnwidth]{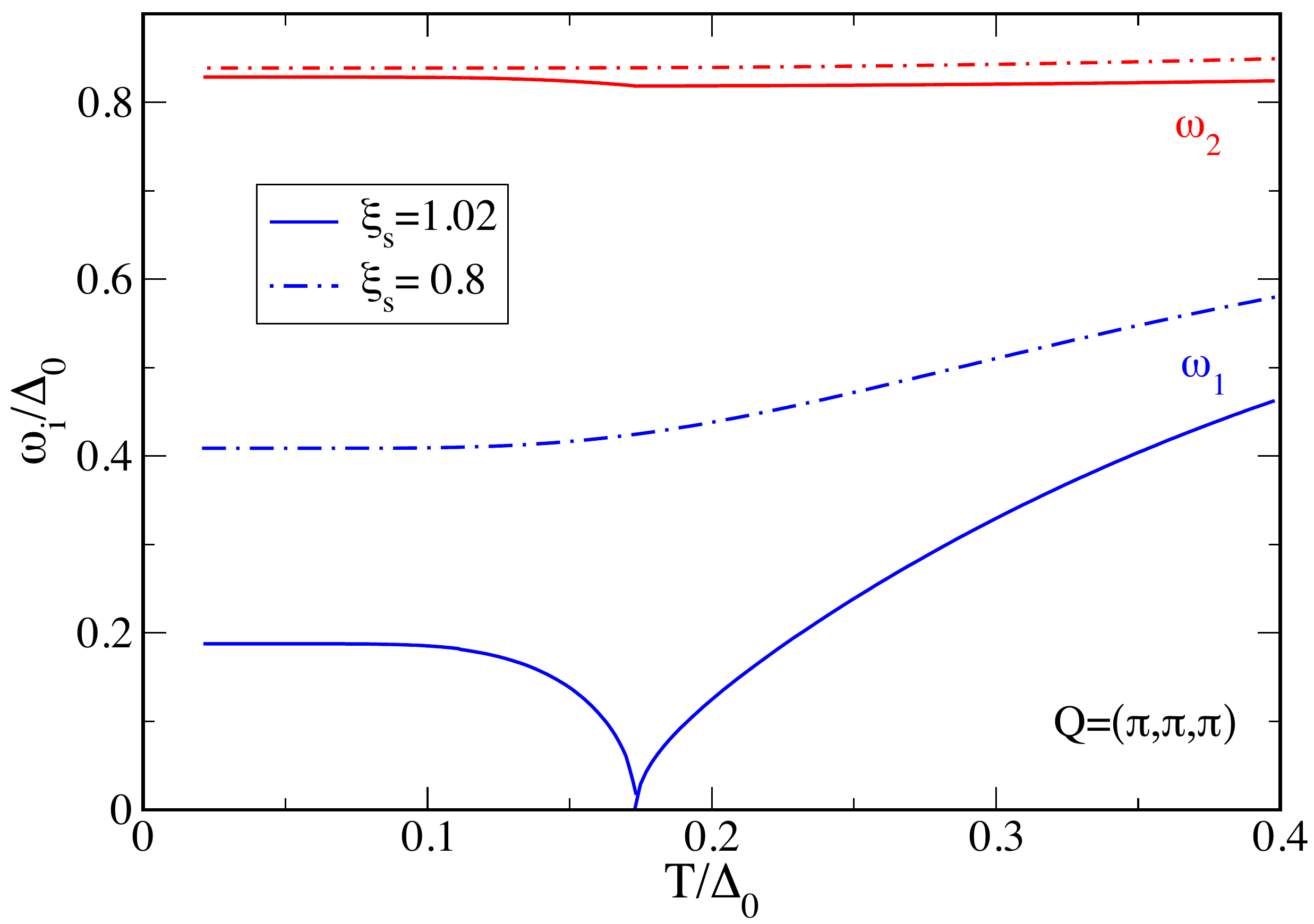}
\caption{Temperature dependence of zone boundary exciton modes $\omega_{1,2}(\bQ)$ for subcritical 
value $\xi_s=0.8$ (broken lines) and above-critical value $\xi_s=1.02$ ($\xi_a=-0.2$ for both). The critical mode $\omega_1$ (connected with the $|1\rangle \leftrightarrow |2\rangle$ transition) softens at $T_m$ but rebounds immediately below. The upper
mode  is hardly affected by AF order (cf. Fig.~\ref{fig:disptemp}), (scheme as Fig.~\ref{fig:TSM3}(a)).}
\label{fig:softmode}
\end{figure}
%%%%%%%%%%%%%%%%%%%%%%fig%%%%%%%%%%%%%%%%%%%%%%%%%%%%%%%
%

\section{The induced order phase and its excitations}
\label{sect:AF}

We now consider the phase with induced magnetic order in the TSM. To be specific we treat only 
the AF case corresponding to the soft mode with at $ \bQ =(\pi,\pi,\pi)$. The more direct treatment 
is based on the RPA approach with the inclusion of the mean field induced order. The alternative would be
the exciton condensation picture for Bogoliubov quasiparticles. The latter is problematic to extrapolate to the 
disordered phase with temperatures considerably above $T_m$ due to the influence of thermally populated 
CEF singlets. This is no problem for the response function approach which will therefore be used here.
As a necessary basis we need the mean field selfconsistency equation for the induced order parameter.
The CEF molecular field Hamiltonian is given by
\bea
H^{mf}_{CEF}&=&H_{CEF}-\sum_l h_e(l)J_z(l)\nonumber\\
h_e(l)&=&\sum_{l'}I_{ll'}\langle J_z(l')\rangle
\label{eqn:mfham}
\eea
with the exchange model of Eq.~(\ref{eqn:iex}) the effective molecular field on the two AF sublattices $A,B$ is $h^{A,B}_e=I_e\langle J^{B,A}_z\rangle$ where $I_e=z|I_0|$ ($I_0 <0$ for AF exchange) and  $\langle J^{A,B}_z\rangle =\pm \langle J_z\rangle $ .  
The associated difference in free energy per site between  induced moment state and paramagnetic state corresponding to Eqs.~(\ref{eqn:mfham},\ref{eqn:CEFham}) is given by
\bea
\delta F/N=-I_e\langle J_z\rangle^2-T\sum_i(p'_i\ln p'_i - p_i\ln p_i)
\eea
where the $p_i$  are the paramagnetic CEF level occupations (Sec. \ref{sec:model}) and $p'_i$ the occupation of levels  in the AF state, renormalized by the molecular field. Explicitly,  $p'_i=Z^{'-1}\exp(-E'_i/T)$  $(Z^{'-1}=\sum_i\exp(-E'_i/T))$. The modified molecular field CEF energies $E'_i$ and eigenstates $|\psi'_i\rangle$ are derived and discussed in Appendix \ref{sect:app1}.

\subsection{Order parameter and saturation moment}
\label{subsect:OP}

Calculating the diagonal (elastic) matrix elements of $J_z$ within the mf eigenstates $|\psi'_i\rangle$  the selfconsistency equation of the order parameter may be given as
\bea
\langle J_z\rangle_T = 2\sum_ip'_i[\beta u_iw_i+v_i(\alpha u_i+\tal w_i)]
\label{eqn:op}
\eea
The primed quantities generally refer to the MF values in the ordered state with nonvanishing $\langle J_z\rangle$. The latter appears implicitly in Eq.(\ref{eqn:op}) through the mf energy levels $E'_i$ and the coefficients $(u_i,v_i,w_i), i=0,1,2$ of the wave functions $|\psi'_i\rangle$. The resulting temperature dependence of the AF order parameter $\langle J_z\rangle$ below $T_m$  together with the paramagnetic inverse static RPA susceptibility $\chi^{-1}(0,T)$ above $T_m$ is shown in Fig.~\ref{fig:OP-sus} for a value of $\xi_a <1$ that results in a second order transition. The divergence of  $\chi(0,T)$  at $T_m$ triggers the appearance of the induced moment $\langle J_z\rangle$. The latter is due to the mixing of excited $|1\rangle, |2\rangle$ CEF states into the mf ground state $|\psi_0\rangle$ (Eq.~(\ref{eqn:mfstate})).
 The figure also displays the T-dependence of selfconsistent admixture coefficients $v_0,w_0$ of excited states $|1\rangle,|2\rangle$ into the molecular field ground state $|\psi'_0\rangle$ according to Eq.~(\ref{eqn:mfstate}).
 In contrast the similar Fig.~\ref{fig:OP-sus1} presents the case of the first order transition $(\xi_a >1)$ for two different $T_m$. There the susceptibility $\chi(0,T)$  no longer diverges at $T_m$ and the order parameter $\langle J_z\rangle$ and admixture coefficients jump to a finite value. From tracing  $\langle J_z\rangle$=0 for different $\xi_s$ the $1^{st}$ order transition line in the inset of  Fig.~\ref{fig:tm1st-inset} (red symbols) may be obtained.\\
 
 The saturation moment at zero temperature is obtained from Eq.~(\ref{eqn:op}) as
\bea
\langle J_z\rangle_{T=0}=2u_0(\alpha v_0+\beta w_0)
\label{eqn:satmom}
\eea
where on the r.h.s the index zero refers to the ground state $|\psi'_0\rangle$. For the TSM this equation cannot be solved explicitly for $\langle J_z\rangle$ since the latter enters on the r.h.s in a complicated manner in the mixing coefficients and associated mf energies (see App. \ref{sect:app1}). As a reference we give the expression for the {\it two-singlet} case (discarding the state  $|2\rangle$ for the moment) where it can be derived \cite{thalmeier:02} explicitly as
\bea
\langle J_z\rangle_{T=0}=\alpha\frac{1}{\xi_\alpha}(\xi_\alpha^2-1)^{\fs} =
\left\{
\begin{array}{rl}
\alpha \; &\xi_\alpha\gg 1\\
\alpha(2\delta)^\fs \;&\xi_\alpha\simeq 1+\delta
\end{array}
\right.
\eea
Thus the saturation moment and its ratio to the transition temperature $(\langle J_z\rangle_{T=0}/\alpha)/(T_m/\Delta)=(2\delta)^\fs\ln(\frac{2}{\delta})\rightarrow 0$ vanish when the induced magnet is close to the quantum critical point, i.e. $\delta\rightarrow 0$.  This is in marked contrast to a conventional semiclassical (degenerate $S=\fs$) magnet \cite{majlis:07}  where the corresponding ratio is constant, given by  $\langle S_z\rangle_{T=0}/(T_m/I_e)=1$ in that case.
This  peculiar dependence of saturation moment  and its ratio with the transition temperature on the control parameters is also apparent in the TSM   (Eq.~(\ref{eqn:satmom})) as presented in Fig.~\ref{fig:satmom}. The saturation moment (now normalized to $m_0=(\alpha^2+\beta^2)^\fs$) increases with square root-like behaviour above the critical parameter $\xi^c_s=1$ approaching unity for $\xi_s\gg 1$. Because $T_m$ varies only logarithmically for $\xi_s\rightarrow \xi^c_s$ the ratio of both quantities (blue line) first increases and the rapidly drops to zero.

\subsection{Collective excitations in the AF phase}
\label{subsect:AFexc}

With $\langle J_z\rangle$ determined we now may compute the renormalized excitation spectrum in RPA approach in the induced moment phase. For this purpose we need the renormalized local CEF energy differences $E'_{ij}=E'_i-E'_j$ of molecular field states  (Eq.(\ref{eqn:mf-level}))   and the inelastic matrix elements between them which lead to renormalized matrix elements $\alpha', \beta' ,\tal'$ 
which are now generally {\it all} non-vanishing because $\Theta$ is broken (Appendix \ref{sect:app1}). In addition we define modified effective temperature dependent parameters
$\alpha'^2_T=\alpha'^2p'_{01}$,  $\beta'^2_T=\beta'^2p'_{02}$,  $\tal'^2_T=\tal'^2p'_{12}$ analogous to  Eq.(\ref{eqn:Tcontrol}).
With the replacements $(\Delta,\Delta_0,\tDe)\rightarrow (E'_{10}, E'_{20}, E'_{21})$ and 
$(\alpha^2_T, \beta^2_T, \tal^2_T)\rightarrow (\alpha'^2_T, \beta'^2_T, \tal'^2_T)$ the exciton mode frequencies in the induced AF ordered phase may be obtained from Eqs.({\ref{eqn:baremode1},\ref{eqn:RPAmode}) by substitution. 

An example of the temperature dependence of the exciton dispersions $\omega_{1,2}$ is presented in Fig.~\ref{fig:disptemp} using the parameter set of Fig.~\ref{fig:OP-sus} ($2^{nd}$ order case) for temperatures above, at and below $T_m$. The flat mode $\omega_3$ in Fig.~\ref{fig:RPA-bogol} which has vanishing intensity is not shown here. The $\omega_1$ dispersion displays the typical softmode behaviour when temperature is lowered down to $T_m$ (dashed, full lines). However, immediately below $T_m$ the dispersion shifts to finite frequency again (dash-dotted).

The corresponding continuous temperature dependence for the $\bQ=(\pi\pi\pi)$ soft mode with the same parameter set and another subcritical one for comparison is shown in Fig.~\ref{fig:softmode}. In the latter  (broken lines)  the zone boundary $\omega_1(\bQ)$ mode softens but then stays flat with lowering temperature while the upper mode $\omega_2(\bQ)$ is practially constant, see also Figs.~\ref{fig:RPA-bogol},\ref{fig:LogSpec3}.
When $\xi_s$ is above critical value (as in Fig.~\ref{fig:disptemp})  $\omega_1(\bQ)$ now actually hits zero, triggering the onset of AF order shown in Fig.~\ref{fig:OP-sus}. Once the molecular field $h_e$ becomes finite and increases the splittings between renormalized levels at $E'_i$ the critical mode  $\omega_1(\bQ)$  is again stabilized to finite frequencies already seen in  Fig.~\ref{fig:disptemp} for $T<T_m$. On the other hand the upper mode for the parameters used shows very little temperature effect. It is certainly possible to fine-tune the parameters such that both modes of the TSM become critical or closely so, but this seems rather artificial and physically one normally has to deal with just one critical mode as is the case e.g. in the cubic singlet-triplet system Pr$_3$Tl\cite{buyers:75}.
From the above disucussion it is clear that the softening of the critical mode in the case of a $1^{st}$ order transition is arrested at a finite energy value.

\section{Summary and conclusion}

In this work we have given a complete survey of a most general extended three-singlet model (TSM) of induced moment quantum magnetism. It consists  of three nonmagnetic CEF singlets coupled by non-diagonal matrix elements of one of the angular momentum components constrained by time reversal $\Theta$. Such low-lying TSM configurations occur frequently in rare earth or actinide compounds with $4f^2$ or $5f^2$ or other even occupation f-electron configurations. The model may be characterized by individual three  $(\xi_\alpha,\xi_\beta,\xi_{\tal})$  but effectively two  $(\xi_s,\xi_a)$ dimensionless control parameters, involving the CEF splittings, non-diagonal matrix elements and intersite exchange.\\

We used two approaches to calculate the elementary excitation spectrum as function of control parameters; the response function RPA formalism and the Bogoliubov quasiparticle approach. They agree on the basic properties of the magnetic exciton dispersions and their soft mode behaviour as function of  $(\xi_s,\xi_a)$. While the latter approach is only practical at low temperature range but gives the Bloch states of exciton bands the RPA formalism covers all temperatures and in particular  the mode softening as function of temperature and the criticality condition
for the onset of induced magnetism as function of  $(\xi_s,\xi_a)$. 

As a new aspect of the TSM we showed that for suitable control parameters a temperature induced hybridization of modes takes place
with an anticrossing of the two exciton dispersions resulting from excitations out of the ground state. On the other hand a possible thermally excited mode stays dispersionless and is only visible at elevated temperatures.

The criticality condition leads to an equation for the dependence of ordering temperature on control parameters which may be solved explicitly for $T_m$ in the weakly and partly symmetric cases. For $\xi_a<1$ the condition for a finite induced ordering temperature is always given by $\xi_s>1$, independent of the values of $\xi_a$ and the splitting ratio r. Furthermore in this case the transition is always
of second order as evidenced by the calculation of paramagnetic susceptibility and temperature dependent order parameter.

Another  possibility not observed in the singlet-singlet case arises when we consider the phase diagram and magnetic ordering temperature $T_m(\xi_s,\xi_a)$  for  $\xi_a>1$ which means that the thermally excited nondiagonal processes are important.
Then the second order transition at $T_m$ extends to control parameter $\xi_s<1$ and finally turns into a first order transition at the 
critical point $T^*$, as is demonstrated by the behaviour of susceptibility and order parameter temperature dependence.

The latter is obtained from the mean-field selfconsistency equations. The resulting molecular field enters into the dynamics via renormallized local CEF energies and non-diagonal matrix elements. Their influence leads to a resurgent stiffening of the soft
mode immediately below $T_m$ which mimics the order parameter. The stiffening is continuous when the transition at $T_m$ is of second
order and has jump-like behaviour for the first order transitions. 

These predicted features may play a role in real Pr- and U- base singlet excitonic magnets and deserve further experimental investigations. This also refers to pressure experiments. The latter may change the CEF splittings and matrix elements and
hence the control parameters and therefore may allow to tune between the different phases found in this three-singlet model 
investigation.

%\section*{ACKNOWLEDGMENTS}
%

\appendix
\section{Example: Three singlet model from tetragonal $(D_{4h})$ $f^2$ $(J=4)$ CEF states}
\label{sect:app0}

 As a concrete example for tetragonal $D_{4h}$ TSM we discuss a TSM level scheme derived from the $J=4$ ninefold degenerate total angular momentum  multiplet  relevant for $4f^2$ and $5f^2$ configurations. In D$_{4h}$ point group CEF environment
 there are five singlets and two doublets \cite{kusunose:11,sundermann:16}. This fact rests solely on the symmetry reduction of total angular momentum representation of the full rotation group corresponding to $J=4$ to the $D_{4h}$ representations  $\Gamma^{(1)}_1,\Gamma^{(2)}_1,\Gamma_2$ (singlet group $s_1(+1)$), $\Gamma_3,\Gamma_4$ (singlet group $s_2(-1)$) and (non-Kramers) doublets $\Gamma_5^{(1)},\Gamma_5^{(2)}(0))$ where the number in parentheses indicates the character of the representation under $C_4$ rotation.
 
 The wave functions and sequences of singlet and doublet energies in a concrete case are then to be obtained, in the simplest manner, by a local CEF Hamiltonian derived from a point charge model (PCM) \cite{hutchings:64,LLW:62} describing the crystal environment and expressed in standard Steven's operator technique. Thereby the PCM parameters are commonly considered as free parameters to be determined from experiment (e.g. temperature dependence of susceptibility and INS peak positions and intensities). The energies $E_i$ and wave functions $|i\rangle$ $(i=1...(2J+1))$ of CEF multiplets are then determined by the five $B_{mn}$ $(mn)=(20),(40),(44),(60),(64))$ CEF parameters for $D_{4h}$ symmetry.
 
 For example in \UR~ it was originally proposed by Santini et al \cite{santini:94} that the lowest states are the three singlets of group $s_1$.  Explicitly they are expressed in terms of $|J_z=M\ket$ free ion states (z refers to the tetragonal axis) as
 \bea
 |\Gamma^{(1)}_1\ket= &&\cos\theta_{c}|0\ket+\sin\theta_{c}\frac{1}{\sqrt{2}}(|4\ket+|-4\ket)\nonumber\\
 |\Gamma^{(2)}_1\ket=-&&\sin\theta_{c}|0\ket+\cos\theta_{c}\frac{1}{\sqrt{2}}(|4\ket+|-4\ket)\nonumber\\
 |\Gamma_2\ket= &&\frac{1}{\sqrt{2}}(|4\ket-|-4\ket)
 \label{eqn:CEFstate}
 \eea
 Within the PCM their energies, referenced to the center of gravity of the three singlets, are given by \cite{kusunose:11}
 \bea
 E(\Gamma^{(1,2)}_1)=\delta(\frac{1}{3}\pm\frac{1}{\cos 2\theta_c});\;\;\;
 %\nonumber\\
% E(\Gamma^{(2)}_1)&=&\delta(\frac{1}{3}-\frac{1}{\cos\theta_c})\nonumber\\
 E(\Gamma_2)=-\frac{2}{3}\delta
 \eea
 where $\delta$ is a splitting parameter and $\theta_c$ a $\Gamma_1$ mixing parameter $(0\leq\theta_c\leq\frac{\pi}{2})$ both determined by the $B_{mn}$. This means within the local CEF-PCM one should have two possible singlet $(s1)$ sequences $|0\ket-|1\ket-|2\ket$ given by   (I) $\Gamma^{(1)}_1-\Gamma_2-\Gamma^{(2)}_1$ or inversely (II) $\Gamma^{(2)}_1-\Gamma_2-\Gamma^{(1)}_1$ depending on the size of $\theta_c$ and the sign of $\delta$. Therefore $\Gamma_2$ should always lie between the two $\Gamma^{(1,2)}_1$ singlets. At the most it can be accidentally degenerate with the lower $(\delta>0)$ or upper $(\delta<0)$ $\Gamma_1^{(1,2)}$ singlet for $\theta_c=0,\frac{\pi}{2}$. Recent NIXS experiments \cite{sundermann:16} advocate that sequence (I) is realized with $\theta_c\simeq\pi/2$. On the other
 hand an alternative sequence with a $\Gamma_2$ ground state and $\Gamma^{(1,2)}_1$ excited singlets different from the simple CEF model has also been proposed from both experiment and DMFT theory \cite{haule:09,kung:15}. In the above TSM only $J_z$ has matrix elements and for sequence (I) they are given by (c.f Fig.~\ref{fig:occ-level})
 \bea
 (\al,\beta,\tal)&=&(4\sin\theta_c,0,4\cos\theta_c)
 \eea
 and $\al,\tal$ interchanged for sequence (II). As in the orthorhombic case (Sec.~\ref{subsect:ortho}) the dipolar matrix element between states with equal time reversal symmetry (here $\Gamma_1^{(1,2)}$ with $\beta=\bra 0 |J_z| 2 \ket$) vanish. Thus staying strictly within the PCM the tetragonal TSM can only support one of the possible dipolar excitation models shown Fig.~\ref{fig:TSM3}(c). Therefore the more flexible lower orthorhombic symmetry which should enable all cases in Fig.~\ref{fig:TSM3} has been chosen in Sec.~\ref{subsect:ortho}.\\

\section{Molecular field energies, states and matrix elements in the induced AF phase}
\label{sect:app1}
In this Appendix we calculate the local mean field energies, eigenstates and matrix elements in the ordered AF phase 
characterized by a selfconsistent order parameter $\langle J_z\rangle=\langle J_z\rangle_A=-\langle J_z\rangle_B$ given by Eq.~(\ref{eqn:op}) where A,B denote the AF sublattices. The total local Hamiltonian of Eq.~(\ref{eqn:mfham}) may be written
explicitly as  
\bea
H^{mf}_{CEF}=
\left(
 \begin{array}{ccc}
-\De&-\alpha_e &-\beta_e\\
 -\alpha_e&0&-\tal_e\\
-\beta_e&-\tal_e& \tDe
\end{array}
\right)\\
\nonumber
\eea
with $\alpha_e=\alpha h_e$, $\beta_e=\beta h_e$ and $\tal_e=\tal h_e$. Here $h_e=I_e\langle J_z\rangle$ 
is the molecular field and we abbreviate $I_e=I_e(\bQ)=-zI_0 >0$ with $\bQ=(\pi,\pi,\pi)$ the AF ordering vector.
The eigenvalues $E'_i(h_e)$ of the molecular field Hamiltionian are then again given by the solutions of the cubic secular equations $(i=0,1,2$):
\bea
E'_i&=&2\bigl(\frac{|p|}{3}\bigr)^\fs\cos\bigl(\frac{\varphi}{3}+\varphi_0\bigr)-\frac{a}{3}\nonumber\\
\varphi&=&\cos^{-1}\Bigl[-\frac{q}{2}\bigl(\frac{|p|}{3}\bigr)^{-\frac{3}{2}}\Bigr]
\label{eqn:mf-level}
\eea
where $\varphi_0=\frac{2\pi}{3}, \varphi_1=\frac{4\pi}{3}, \varphi_2=0$ and $p=\frac{1}{3}(3b-a^2)$; $q=\frac{2}{27}a^3-\frac{1}{3}ab+c$; with the cubic secular equation parameters defined by
\bea
a&=&\De-\tDe\nonumber\\
b&=&-[\De\tDe+(\alpha_e^2+\tal_e^2+\beta_e^2)]\nonumber\\
c&=&\alpha_e^2\tDe^2-\tal_e^2\De+2\alpha_e\tal_e\beta_e
\eea
We formally keep the last term in c although it must vanish identically because one of the matrix elements has to be equal to zero due to time reversal symmetry (Sec.~\ref{subsect:ortho}).
The phases $\varphi_i$ are denoted such that for the paramagnetic case with $h_e=0$ we recover $E'_i=\hat{E}_i=-\De,0,\tDe$ for $i=0,1,2$ consecutively, corresponding to the sequence in Fig.~\ref{fig:occ-level}.
The  associated molecular field orthornormal eigenvectors  are
\bea
|\psi'_i\rangle =u_i|0\ra +v_i|1\ra +w_i|2\ra
\label{eqn:mfstate}
\eea
These coefficients may be obtained for the general model by elimation from the eigenvalue equation $H^{mf}_{CEF}|\psi'_i\rangle = E'_i|\psi'_i\rangle$. It is convenient to introduce the auxiliary factors
\bea
\rho_i=\frac{\alpha_e(\tDe-E'_i)+\tal_e\beta_e}{E'_i(\tDe-E'_i)+\tal_e^2};\;\;
\tilde{\rho}_i=\frac{\beta_eE'_i-\tal_e\alpha_e}{E'_i(\tDe-E'_i)+\tal_e^2}\nonumber\\
\eea
Then the coefficients of MF eigenfunctions are given by 
\bea
u_i=(1+\rho^2+\tilde{\rho}^2)^{-\fs};\; 
v_i=-\rho_iu_i;\;
w_i=\tilde{\rho}_iu_i\nonumber\\
\eea
For the nondiagnonal matrix elements $\langle \psi_i|J_z|\psi_j\rangle$ $(i\neq j)$
we obtain
\bea
\alpha'&=&\alpha(u_0v_1+v_0u_1)+\beta(u_0w_1+w_0u_1)+\tal(v_0w_1+w_0v_1)      \nonumber\\
\beta'&=&\alpha(u_0v_2+v_0u_2)+\beta(u_0w_2+w_0u_2)+\tal(v_0w_2+w_0v_2)\nonumber\\
\tal'&=&\alpha(u_1v_2+v_1u_2)+\beta(u_1w_2+w_1u_2)+\tal(v_1w_2+w_1v_2)\nonumber\\
\label{eqn:matmag}
\eea
We note that in he ordered state with $\Theta$-symmetry broken all $(\alpha',\beta',\tal')$ are nonzero due to the mixing of singlets by the molecular field, although one element of the set $(\alpha,\beta,\tal)$ must always vanish due to $\Theta$-symmetry in the nonmagnetic case. In the discussion of numerical results in Sec.~\ref{sect:AF} we restrict to the case $\alpha =0$  
corresponding e.g. to Fig.~\ref{fig:TSM3}(b).  From these matrix elements the renormalized effective T-dependent matrix elements  $(\alpha'^2_T, \beta'^2_T, \tal'^2_T)$ (see Sec.~\ref{sect:AF}) which appear in the dispersions $\omega'_i(\bq)$ and spectral function of $\chi_0(\bq,\om)$ in the AF ordered phase may be calculated in analogy to  Eq.(\ref{eqn:Tcontrol}).

%%%%%%%%%%%%%%%%%%%%%%%%%      References        %%%%%%%%%%%%%%%%%%%%
%\newpage
%\bibliographystyle{prsty}
\bibliography{References}

\end{document}